\documentstyle[aps,epsfig]{revtex}

\begin{document}
\draft

\title{Quantum Chaos and Thermalization for Interacting Particles
\thanks{to be published in Proceedings of the International School of
Physics "Enrico Fermi", Varenna, Italy, 1999}
}

\author{
F.M.Izrailev
}

\address{
Instituto de F\'isica, Universidad Autonoma de Puebla, Apdo. Postal J-48,
Puebla, 72570 M\'exico
}

\date{\today}
\maketitle

\begin{abstract}

In this review the problem of statistical description of isolated quantum
systems of interacting particles is discussed. Main attention is paid to a
recently developed approach which is based on chaotic properties of compound
states in the basis of non-interacting particles. In order to demonstrate
the most important aspects of this approach, the matrix model of two-body
random interaction between Fermi-particles has been used. Different problems
have been considered such as the onset of chaos and statistical equilibrium,
the relation between the structure of eigenstates and distribution of
occupation numbers, the emergence of the Fermi-Dirac distribution in
isolated systems of finite number of particles and many others. The
application of the approach to dynamical systems with the classical limit is
discussed as well. 
\end{abstract}


\section{Introduction}

Until recently, the {\it quantum chaos theory} was mainly related to
few-body physics. On the other hand, in real physical systems such as
many-electron atoms and heavy nuclei, the origin of complex behavior is
quite strong interaction between many particles. To deal with such systems,
famous statistical approach has been developed which is based on the {\it %
Random\ Matrix Theory} (RMT) (see, for example, 
\cite{P65,M67,brody,GMW99}). The main idea
of this approach is to forget about a detailed description of the motion and
to treat these systems statistically having in mind that the interaction
between particles is so complex and strong that generic properties are
expected to emerge. Simplest models of the RMT are full random matrices of a
given symmetry, the latter was shown to have a direct link with underlying
symmetries of physical systems.

One of the main results of the RMT is the prediction of a specific kind of
correlations in the energy spectra of complex quantum systems. Among many
characteristics of these correlations, the most popular is the distribution
of spacings between nearest energy levels in the spectra. Exact analytical
expression of this distribution is very complicated, instead, one uses the
so-called Wigner-Dyson (WD) surmise (quite simple expression which gives a
very good approximation to the exact result). A distinctive property of this 
{\it WD-distribution} is the repulsion between neighboring levels in the
spectra, the degree of this repulsion (linear, quadratic or quartic) depends
on the symmetry of random matrices. In fact, such type of repulsion was
observed experimentally very long ago (first experimental
observation is reported in Ref.\cite{G39} for the energy spectra of heavy
nuclei), and discussed in many theoretical works.

After this prediction of the RMT, the WD-distribution has been confirmed to
occur in heavy nuclei and many-electron atoms, see references in \cite
{brody,BG84}. Later on, it was found also in dynamical systems with chaotic
behavior in the classical limit, famous examples are the so-called billiards
(see for example, \cite{BG84}). As a result, it was understood that chaotic
properties of quantum systems are generic for both disordered models (when
the randomness of matrix elements is postulated from the beginning) and
dynamical systems for which the pseudo-randomness appears as a result of
special conditions, the latter are convenient to explain by comparing with
the classical limit. Thus, one can say that limiting properties of quantum
chaos in the case when all regular dynamical
effects are neglected, are described by the RMT.

As one can see, the RMT can give a proper description of a system (mainly,
the properties of energy spectra) only locally, in a restricted region of
energy spectra. Indeed, the RMT can not give any global energy dependence
neither for the energy spectrum nor for eigenstates, it is
parameter-independent theory. In this sense, the conventional RMT (ensembles
of fully random matrices) is very restricted and,
for example, it can not be directly applied to
such important phenomena as the localization of eigenstates in
disordered models. This is why new approaches in the RMT have been developed
by imposing internal structure of random matrices. The most known example is
the so-called {\it Band Random Matrices} (BRM), or random matrices with a
band-like structure (see, for example, \cite{I90}, \cite{FM94} and
references therein). Inside the band, the matrix elements are assumed to be
random and independent, and outside the band matrix elements are set to
zero. The BRM-ensemble has been recently studied in details both numerically
and theoretically, and much is now known about the structure of eigenstates and
spectrum statistics for both infinite and finite matrices. The application
of this kind of matrices in physics is very broad. In particular, they have
been used to describe dynamical localization in dynamical systems with
time-periodic perturbation (paradigmatic model is the Kicked Rotor \cite{I90}%
), and the localization of eigenstates in quasi-1d disordered models in
solid state physics \cite{FM94}.

Another class of band random matrices has been introduced very long ago by
Wigner in Ref.\cite{W55}. The structure of these {\it Wigner Band Random
Matrices} (WBRM) is characterized by the {\it leading diagonal} with
reordered (in an increasing way) values, plus random and independent
off-diagonal elements inside the band of size $b\,.$ This type of matrices
is much closer to physical realistic systems, compared to the standard full
random matrices. One can suggest that the original motivation of Wigner for
the study of these matrices was a close correspondence to Hamiltonians of
complex nuclei, which are typically described by the mean-field part $H_0\,$
(leading diagonal) and the residual interaction $V$ of the finite energy
range (off-diagonal matrix elements inside the band). The main interest of
Wigner was the quantity which nowadays is known as the {\it strength function%
} or {\it local density of states} (LDOS). This quantity is extremely
important when describing the spread of energy, initially concentrated in a
specific state of the unperturbed Hamiltonian $H_0\,$, between all other
states due to the internal interaction $V$ . In particular, it was
analytically shown that this function has the form of the Lorentzian if 
interaction is sufficiently (but not extremely) large. This result is
fundamental and it is used in very different applications.

However, in 
spite of the success of the standard RMT\ and its modern developments,
the strong assumption of randomness of matrix elements, as well as a
specific (band) structure of matrices do not allow to relate such matrices
directly to realistic many-particle Hamiltonians. One of the important
reasons is that the underlying structure of realistic Hamiltonian matrices
results from the single-particle spectrum and 
{\it two-body interaction} between
particles. In order to understand the role of $k-$body (random) interaction,
a new ensemble of matrices has been suggested (see, for example, 
\cite{FW70,BF71,brody} and references therein
). In this approach the matrices arise as a result of construction from the
single-particle basis, provided random character of the $k-$body
interaction. The study of this ensemble of matrices have shown that even
when the interaction $V$ is very strong and the influence of the leading
diagonal may be neglected, there are serious differences from the standard
RMT. In particular, it was discovered that for the two-body interaction, $%
k=2 $ , the spectral fluctuations are different from those predicted by the
RMT, although the distribution of spacing between nearest levels has the
form similar to the WD-distribution. It was shown that full random matrices
occur when the rank of interaction is very large, $k\rightarrow \infty \,.$

Due to very serious mathematical problems these {\it two-body random
interaction} (TBRI) matrices were forgotten for quite a long time, and only
recently they have been used in the context of quantum chaos. In these
lectures, the author gives a review of recent results obtained for the
TBRI-model in collaboration with the co-authors. Main attention is paid to a
novel approach developed in \cite{FIC96,FGI96,FI97a,FI97,FI99}, 
which is based on the chaotic
structure of eigenstates in a given basis of unperturbed many-particle
states. This approach allows to relate statistical properties of exact
eigenstates in many-body representation directly to properties of
single-particle operators, in the first line, to the occupation number
distribution of single-particle states.

The structure of the paper is as follows. In the next Section 2.1 the
structure of the TBRI-matrices is discussed in details. It was explained how
these matrices are constructed from single-particle states and what are
properties of matrix elements in many-body representation, also, the
comparison is made with full random matrices. A particular point is that in
spite of complete randomness and independence of two-body matrix elements,
off-diagonal matrix elements of many-particle
Hamiltonian matrices have underlying
correlations which are due to a two-body nature of interaction.

In Section 2.2 generic properties of density of states and level spacing
distribution are briefly discussed, although they are not the main interest
of the study. Section 2.3 deals with the structure of exact eigenstates in
dependence on the interaction strength and excitation energy. The notion of
the average shape of the eigenstates ($F-${\it function}) is introduced and
discussed in details since the structure of chaotic eigenstates plays a basic
role in the approach. In next Section 2.4 the structure of the strength
function is considered and compared with that of exact eigenstates. Main
attention is paid to the conditions under which this form is the Lorentzian,
and how this form changes with an increase of interaction strength. In
Section 2.5 very recent analytical results are discussed, which are obtained
for the shape of the strength function in the TBRI-model for any strength of
interaction. Section 2.6 is devoted to non-statistical properties of this
model. Specifically, it is shown that some quantities can not be described
in a statistical way, in spite of completely random character of the
two-body interaction.

Next Section 3.1 starts with the discussion of the basic relation between
the structure of exact eigenstates and the 
{\it distribution of occupation numbers } (DON). It is shown that the average
shape of eigenstates plays a crucial role for this distribution and
practically determines mean values of
single-particle operators. In next Section 3.2 the relevance of the DON for
isolated systems, to the standard canonical distribution is discussed. It is
shown how the canonical distribution emerges in isolated systems with an
increase of the number of interacting particles. In next Section 3.3 the
problem of the Fermi-Dirac distribution is considered for the TBRI-model of
interacting Fermi-particles, in particular, conditions under which this
distribution occurs in isolated systems are analyzed. An important problem
of the statistical description of the occupation number distribution is
considered in Section 3.4. An analytical approach has been developed in
order to obtain the DON in the case of statistical equilibrium which results
from the chaotic structure of eigenstates. This approach is valid even for
small number of particles, in the case when the DON differs from the
Fermi-Dirac distribution. In next Section 3.5 another approach is suggested
for the description of the DON, in the case when its form is of the
Fermi-Dirac type. It was shown how the DON can be obtained by a proper
renormalization of the total energy of a system, originated from the
interaction between particles. General discussion of the meaning of
temperature in isolated systems of finite number of particles is the content
of Section 3.6. The main point is that for small number of particles
different definitions of temperature give different results. Therefore, is
it of interest to compare these definitions and to understand their
meanings, if any. Finally, in Section 3.7 main results are summarized for
the transition to chaos and equilibrium in the TBRI-model in dependence on
the interaction strength. In last Section 4 it is briefly shown how the
developed approach can be applied to dynamical systems.

\section{ Two-body random interaction model}

\subsection{Description of the model}

\subsubsection{Many-body Hamiltonian}

The model we discuss here deals with Hamiltonians which can be separated in
two parts, 
\begin{equation}
\label{H}H=H_0+V 
\end{equation}
where $H_0$ describes the ``unperturbed '' part and $V$ stands for the
interaction between particles or between different degrees of freedom. In
order to study statistical properties of such models we assume in the
following that the interaction is completely random. In contrast with standard
approach of the RMT where matrix elements of $V$ are taken as random
variables, we would like to keep an important physical property of real
systems and to take into account that the interaction is of the two-body
nature. Therefore, we start with the single-particle Hamiltonian which
refers $n$ non-interacting particles
occupying $m$ single-particles levels, and assume that the matrix elements of
the {\it two-body interaction} $V_{s_1s_2s_3s_4}$ are independent random
variables. Here, the indices $s_1,s_2,s_3,s_4$ indicate initial $(s_1,s_3)$
and final $(s_2,s_4)$ single-particle states coupled by the interaction.

In what follows we consider Fermi-particles, however, the approach can be
easily extended to Bose-particles \cite{BGI98,BGIC98}. Therefore, in the
Slater determinant basis the unperturbed part has simple form 
\begin{equation}
\label{H0}H_0=\sum \epsilon _s\,a_s^{\dagger }a_s 
\end{equation}
and the perturbation can be represented as 
\begin{equation}
\label{V}V=\frac 12\sum V_{s_1s_2s_3s_4}\,a_{s_1}^{\dagger }a_{s_2}^{\dagger
}a_{s_3}a_{s_4}. 
\end{equation}
Here $\epsilon _s$ is the energy of a particle, corresponding to the
single-particle state $\left| s\right\rangle $ and $a_{s_j}^{\dagger }$ , $%
a_{s_j}$are creation-annihilation operators. With these notations, exact
eigenstates $\left| i\right\rangle $ of the total Hamiltonian $H$ ({\it %
compound states) }can be expressed in terms of eigenstates $k$ of the
unperturbed part $H_0$ ({\it basis states) }as follows

{\it 
\begin{equation}
\label{basis}\left| i\right\rangle =\sum_kC_k^{(i)}\left| k\right\rangle
,\,\,\,\,\,\,\,\,\,\,\,\left| k\right\rangle =a_{s_1}^{\dagger
}\,.\,\,.\,\,.\,a_{s_n}^{\dagger }\left| 0\right\rangle 
\end{equation}
}where {\it $C_k^{(i)}$} is the $k-th$ component of the compound state {\it $%
\left| i\right\rangle $} in the unperturbed basis. 
These components determine the important quantity which will be discussed in
great details below, 
the {\it %
occupation numbers }$n_s$ ,

\begin{equation}
\label{ns}n_s=\left\langle i\right| \hat n_s\left| i\right\rangle
=\sum_k\left| C_k^{(i)}\right| ^2\left\langle k\right| \hat n_s\left|
k\right\rangle 
\end{equation}
with $\hat n_s=a_s^{\dagger }a_s$ as the occupation number operator. For
Fermi-particles the occupation number $n_s^{(k)}=\left\langle k\right| \hat
n_s\left| k\right\rangle $ is equal to $1\,$or $0$ depending on whether any
of the particles in the basis state $\left| k\right\rangle $ occupies or not
the single-particle state $\left| s\right\rangle $ .

As one can see, our model is described in terms of many-particles basis
states $\left| k\right\rangle \,$ and exact (compound) states $\left|
i\right\rangle $ which are constructed from the single-particle states $%
\left| s\right\rangle $ and two-body matrix elements $V_{s_1s_2s_3s_4}$ . In
what follows we assume that the basis states are reordered in an increasing
way for the total energy $E_k=\sum_s\epsilon _sn_s^{(k)}$ with an increase
of the index $k=1,\,.\,.\,.\,,\,N\,$ . This way of the ordering of the
unperturbed basis is crucial for the analytical description of chaotic
compound states which are formed by the interaction between many basis
states, see below. The size $N$ of the basis for the many-particles
Hamiltonian $H$ can be found from the combinatorics: if any single-particle
state can be occupied by one particle only, one can get 
\begin{equation}
\label{Nsize}N=\frac{m!}{n!(m-n)!}\sim \exp \left( n\ln \,n\,+(m-n)\ln
\,\frac m{m-n}\right) 
\end{equation}
where $m$ is the number of single-particle states ({\it orbitals}). The
latter estimate in (\ref{Nsize}) shows that total number of many-particle
states increases very fast (exponentially) with an increase of the number of
particles and orbitals. For example, for $m=11$ and $n=4$ the size of the
$H-$matrix is $N=330$ . The model with these parameters has been studied in
great details in \cite{FIC96,FGI96,FI97a,FI97} and compared with direct
computations of the Ce atom \cite{FGGK94,GFG95,FGGP99}. 
For this atom there are $n=4$
valence electrons, and the core can be effectively described by the
Hartree-Fock method. 
This method has been used \cite{FGGK94,GFG95,FGGP99} in order to
calculate the basis set of single-particle relativistic states with energies 
$\epsilon _s$ as well as the matrix elements $V_{s_1s_2s_3s_4}$ of 
interaction between valence electrons. The Ce atom is known to have good
statistical properties, and this was the reason in \cite
{FIC96,FGI96,FI97a,FI97} to compare direct calculations with the simplest
model (\ref{H},\ref{H0},\ref{V}) we are going to discuss. In spite of the
fact that this model does not take into account the momentum (it depends
only on the energy and for this reason can be treated as ``zero-dimensional
'', it turns out to be instructive for the comparison of the real ({\it %
dynamical}) Ce atom with the two-body random interaction model ({\it %
TBRI-model}) described above (for details see \cite{FIC96,FGI96,FI97a,FI97}%
). In what follows, for the single-particle energies $\epsilon _s$ we take a
non-degenerate spectrum with constant mean level spacing $d_0=\,<\epsilon
_{s+1}-\epsilon _s>$ which, without the loss of generality can be taken $%
d_0=1$ . The unperturbed single-particle spectrum has been chosen at random,
or according to the expression $\epsilon _s=d_0(1+1/s)$ (the
results are statistically the same). As one can see, the model is defined by
four parameters, $m,\,n,\,d_0$ and $V_0^{\,2}=\left\langle
V_{s_1s_2s_3s_4}^{\,\,\,2}\right\rangle $ which is the variance of two-body
random matrix elements (we assume that the distribution of these elements is
the Gaussian with the zero mean).

The Hamiltonian (\ref{H}) with (\ref{H0}) and (\ref{V}) is of general form,
it appears in many physical applications such as when describing complex
atoms, nuclei, atomic clusters etc. In fact, the form of $H$ discussed above
is known as the {\it mean field approximation} for complex quantum systems
of interacting particles. In this description, the unperturbed part $H_0$
represents the zero-order mean field for the excited states with the ground
state $E_1$, and the {\it residual }two-body interaction is given by $V$ .
Therefore, the single-particle levels $\epsilon _s$ in such applications
are, in fact, renormalized quasi-particle energies (see details, for
example, in \cite{F96}). The considered here model does not take into
account such physical effects as momentum dependence, pairing effects and
others, however, it contains the main effects of quantum chaos and is very
effective for the understanding generic features of complex systems.
Moreover, as was pointed out, the
approach we discuss below, can be extended for {\it dynamical} quantum systems
which exhibits {\it complex behavior}.

From the view point of dynamical systems when the complexity of the behavior
appears as a result of {\it dynamical chaos} (both for the systems with or
without {\it the classical limit}), the separation of the total Hamiltonian
in two parts is well defined physical procedure. Specifically, the approach
is expected to be valid if the second part $V$ is as ``random '' as
possible. In other words, one should find such separation that the
perturbation has no strong regular part, the latter should be embedded into
the ``unperturbed '' Hamiltonian $H_0$ . In this way, the compound states
may be treated, for sufficiently strong perturbation, as chaotic
superposition of simple basis states. As is well known, such a situation is
typical for many-electron atoms and heavy nuclei. Indeed, the number of
basis states (number $N_{pc}$ of {\it principal components }) is known to be
about $10^4-10^6$ for excited nuclei, and $\sim \,100$ in excited rare-earth
or actinide atoms.

By assuming the complete randomness of two-body elements in our models,
we avoid the influence of any regular effects. Therefore, our goal is to
explore statistical properties of the model in its strongest ``chaotic ''
limit. One should stress that the answer is far from being trivial since the
many-body Hamiltonian $H$ turns out to be quite different from standard
random matrix ensembles of the RMT and, as will be shown, some important
quantities can not be described statistically.

It should be noted that the TBRI-model we discuss here, for the first time
was analyzed long ago (see \cite{FW70,BF71}, also, the review \cite
{brody} and references therein). The original interest was related to the
fact that for standard random matrices the density of states has the famous 
{\it semicircle} form, in contrast with physical systems for which the RMT
was addressed (complex nuclei). Moreover, the density of states for
many-body systems increases very fast, and has nothing to do with the
semicircle even on a local scale. Therefore, the natural question is
to understand what is  
missed in the RMT. For this reason, another ensemble of 
matrices has been suggested which takes into account $k-$body interaction
between particles. The theoretical study has been shown that, indeed, fully
random matrices of the kind considered in the standard RMT, formally
correspond to the case $k\rightarrow \infty $ . On the other hand, with
decrease of $k$ , the density of states tends to the Gaussian form. The
latter form is closer to reality and on a local scale, the density
may be treated as the realistic one. Another question which has been under
close investigation, is the role of $k$-body interaction on the spectrum
statistics, see discussion below. For some reason, these studies have not
been extended until recently, when the role of two-body interaction in
different applications was questioned
in the context of quantum chaos. Unlike the previous studies,
below we pay the main attention to chaotic properties of eigenstates, and to
the problem of how these properties can be linked to the properties of
single-particle operators, such as the {\it occupation number} {\it %
distribution}.

\subsubsection{Structure of the Hamiltonian matrix}

Let us start with the structure of the Hamiltonian $H$ in the chosen basis.
For this, one should construct matrix elements $H_{ij}=\left\langle i\right|
H\left| j\right\rangle $ which correspond to the coupling between basis
states $\left| i\right\rangle $ and $\left| j\right\rangle $ due to the
interaction $V$. One can immediately see that the number $N_2=m^2(m-1)^2/2\,$%
of independent matrix elements $V_{s_1s_2s_3s_4}$ of the two-body
interaction is much less than the total number $N(N+1)/2$ of the (symmetric)
matrix elements $H_{ij}$ . It is very important that due to a two-body
character of the interaction, the matrix elements $H_{ij}$ are non-zero only
when basis states $\left| i\right\rangle $ and $\left| j\right\rangle \,$
differ by no more than two occupied single-particle states. In order to
count the total number $K$ of non-zero matrix elements $H_{ij}$ for the
fixed $i$, we separately count the numbers $K_0\,,\,K_1\,,\,K_2$ of non-zero
matrix elements which correspond to the transition between the basis states
which differ by the positions of none, one and two particles, respectively, 
\begin{equation}
\label{K123}K_0=1,\,\,\,\,\,\,\,\,K_1=n(m-n),\,\,\,\,\,K_2=\frac
14n(n-1)(m-n)(m-n-1). 
\end{equation}
As a result, the total number $K$ in each line of the matrix is 
\begin{equation}
\label{Ktot}K=K_0+K_1+K_2=1+n(m-n)+\frac 14n(n-1)(m-n)(m-n-1)\approx \frac
14n^2m^2 
\end{equation}
where the last estimate is given for large number of particles and orbitals, 
$1\ll n\ll m$ . Comparing $K$ with $N\,$ defined by Eq.(\ref{Nsize}) , one
can see that the matrix $H$ is {\it sparse} (only for $n=2$ there is no
forbidden transitions and the matrix is full). 

\begin{figure}[htb]
\vspace{0.0cm}
\begin{center}
\epsfig{file=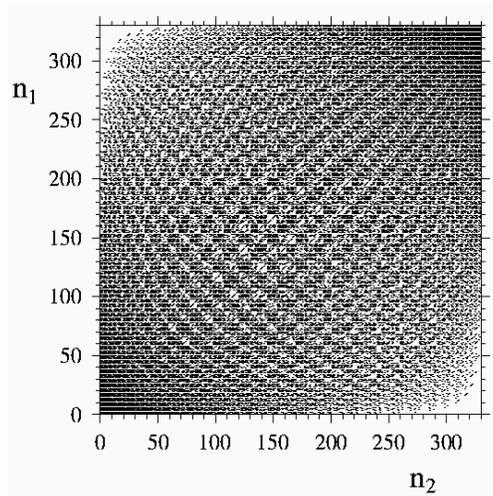,width=2.6in,height=2.6in,angle=-90}
\vspace{0.5cm}
\caption{
Sparsity of the Hamiltonian matrix $H_{n_1,n_2}$ for $n=4$ particles, $%
m=11$ orbitals. Black points are non-zero matrix elements.
}
\end{center}
\end{figure}

The presence of many zeros in
the matrix means that, in a sense, there are strong correlations between
matrix elements since the position of zeros are fixed for any random choice
of the two-body random matrix elements. It should by pointed out that the
sparsity increases with an increase of the number of particles, therefore,
the more particles (and orbitals), the less relative number of
non-zero elements. This fact is important for the description of such
systems by the random matrix approach. One should remind that in
the standard RMT, the {\it sparsity} is not taken into account at all.

The sparsity of the matrix for $m=11$ and $n=4$ is shown in Fig.1 where
black points correspond to non-zero elements. First, one can see that the
density of zero elements increases when moving away from the principal
diagonal. Second, the positions of non-zero matrix elements are correlated,
there are some curves along which the density is high, this reflects the
two-body nature of interaction. In contrast with full random matrices of the
standard RMT, the influence of the off-diagonal elements depends on the
distance from the principal diagonal. To illustrate this peculiarity, we
averaged the modulo of the off-diagonal matrix elements over blocks of the
fixed size $10\times 10$ in such a way that instead of the matrix of the
size $N\times N$ we have the reduced size $N/10\,\,\times \,N/10$ . The
result is shown in Fig.2 where only off-diagonal terms are presented. As one
can see, the amplitude of these effective matrix elements
decreases when moving away
from the diagonal. This means that the effective intensity of the
off-diagonal terms far from the diagonal is less than of those close to the
diagonal. In some sense, one can treat the structure of the Hamiltonian as
the {\it band-like}, although it is clear that the amplitude of the averaged
matrix elements decays quite slowly.

\begin{figure}[htb]
\vspace{-0.5cm}
\begin{center}
\epsfig{file=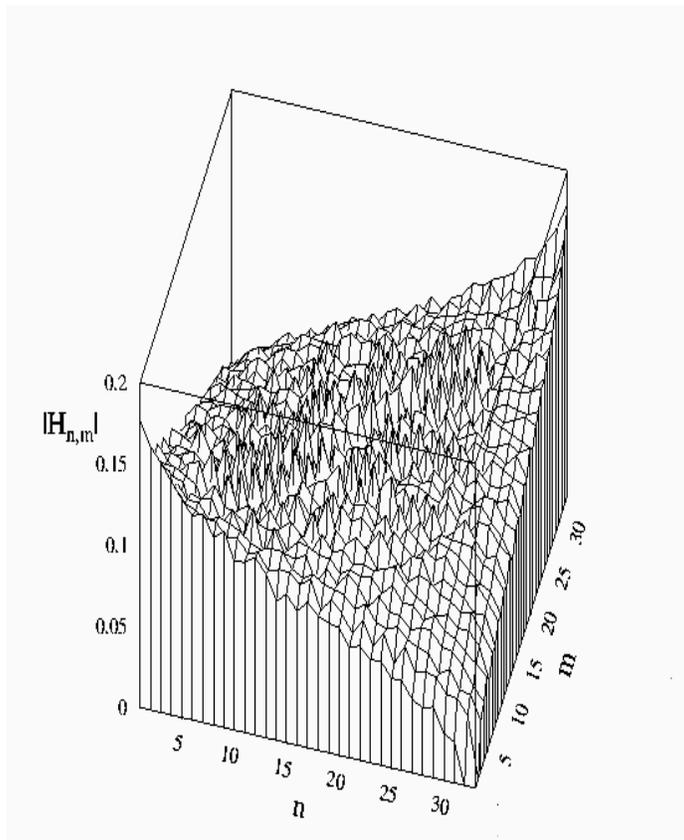,width=4.4in,height=3.6in,angle=-90}
\vspace{0.5cm}
\caption{
Shape of the Hamiltonian matrix (without leading diagonal $H_{i,i}$).
The matrix for $n=4, m=11, V_0=0.12, d_0=1.0$ has been divided into blocks
of size $10\times 10$ and the sum $\left| H_{n,m}\right| =\sum_{i,j}\left|
H_{i,j}\right| \,$ has been computed inside each block $(n,m)$ . This type
of the average allows to see effective intensity of the off-diagonal matrix
elements.
}
\end{center}
\end{figure}

\subsubsection{Correlations in off-diagonal matrix elements}

Now, let us analyze non-zero off-diagonal elements of the matrix $H_{ij}$ .
According to the definition (\ref{V}), any of these elements is just a sum
of one or more two-body matrix elements. When the basis states $\left|
i\right\rangle \,$ and $\left| j\right\rangle $ differ by one occupied
orbital, $\left| j\right\rangle =a_{s_2}^{\dagger }a_{s_1}\left|
i\right\rangle $ , the matrix element $H_{ij}$ is the sum of $n-1$ two-body
matrix elements, 
\begin{equation}
\label{corr1}H_{ij}=\sum_\mu ^{n-1}\,V_{s_1\mu \,s_2\mu }\,=\sum_{\mu \neq
\alpha }^{n-2}\,V_{s_1\mu \,s_2\mu }\,\,+\,V_{s_1\alpha \,s_2\alpha } 
\end{equation}
Here the last equality is given in order to show that among other matrix
elements $H_{i^{\prime }j^{\prime }}$ there are such elements which differ
from $H_{ij}$ by the last term only and the sum of $n-2\,$is exactly the
same \cite{FGI96}, 
\begin{equation}
\label{corr1a}H_{i^{\prime }j^{\prime }}=\sum_\mu ^{n-1}\,V_{s_1\mu \,s_2\mu
}\,=\sum_{\mu \neq \alpha }^{n-2}\,V_{s_1\mu \,s_2\mu }\,\,+\,V_{s_1\,\beta
\,s_2\beta } 
\end{equation}
with $\alpha \neq \beta $ . This happens for those basis states whose
many-particle energies differ by the energy difference corresponding to the
move one particle from the orbital $\alpha $ to the orbital $\beta $ , 
\begin{equation}
\label{Ediff}E_{i^{\prime }}\,-E_i=E_{j^{\prime }}\,-E_j=\epsilon _\beta
\,-\epsilon _\alpha 
\end{equation}
One can see that these matrix elements, strictly speaking, can not be
treated as completely independent variables. Therefore, if one averages over
the ensemble of matrices $H_{ij}$ which constructed from different sets of
two-body random elements, the correlations for such elements remain, $%
\left\langle H_{ij}H_{i^{\prime }j^{\prime }}\right\rangle \neq 0$ .

The more striking result arises when considering matrix elements $H_{ij}$
which correspond to the coupling between those basis states $\left|
i\right\rangle \,$ and $\left| j\right\rangle $ which differ by two occupied
orbitals, $\left| j\right\rangle =a_{s_2}^{\dagger }a_{s_1}\left|
i\right\rangle $ . These matrix elements are equal to the corresponding
single two-body matrix elements, in other words, there is only one term in
the sum in Eq. (\ref{V}), $H_{ij}=\,V_{s_1\mu _1\,s_2\mu _2}$ . These matrix
elements correspond to the move of
two particles from the orbitals $s_1\,,\,s_2$
to other orbitals $\mu _1\,,\,\mu _2\,$ . At the same time, the rest of
particles ( $n-2\,\,$ particles) can occupy different $m-4$ orbitals.
Therefore, the same matrix element stands for other basis states $\left|
i^{\prime }\right\rangle \,$ and $\left| j^{\prime }\right\rangle $ with the
same move of two particles, thus, $H_{i^{\prime }j^{\prime
}}=H_{ij}=\,V_{s_1\mu _1\,s_2\mu _2}$ . As a result, among matrix elements
of the Hamiltonian matrix $H_{ij}$ there are equal matrix elements,
although they are chosen randomly from the ensemble of two-body random
matrices. This non-trivial fact indicates that, in spite of completely
random character of the two-body interaction, the (non-zero) matrix elements
of the many-body Hamiltonian are not completely independent variables!

\begin{figure}[htb]
\vspace{-2.0cm}
\begin{center}
\hspace{-4cm}
\epsfig{file=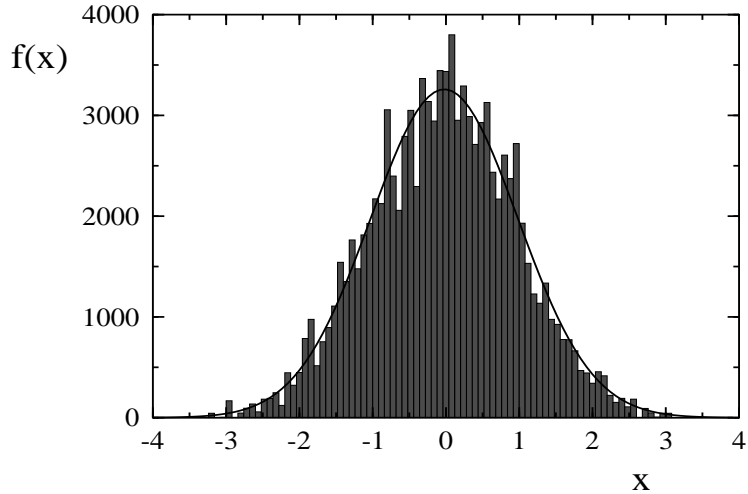,width=4.4in,height=3.6in,angle=-90}
\vspace{-2.0cm}
\caption{
Distribution of normalized off-diagonal matrix elements of the matrix $%
H_{i,j}$ for the parameters of Figs.1-2. Each of matrix element $H_{i,j}$
has been normalized to its variance computed from the average over $N_g=16$
matrices with different realizations of the (random) two-body matrix
elements. The smooth curve is the best fit to the Gauss which is expected
for uncorrelated matrix elements.
}
\end{center}
\end{figure}

The role of the above underlying correlations which are due to a two-body
character of the interaction, is an interesting and important problem (see 
\cite{FGI96}), we will discuss some results in Section 2.6. Here, we would
like to show that these correlations can be easily detected by the study of
the distribution of matrix elements $H_{ij}$ . For this, let us take the
ensemble of the Hamiltonians $H\,$ with different two-body random matrix
elements $V_{s_1s_2\,s_3s_4}$ keeping all other parameters. Then, for any
fixed values $i$ and $j$ , we can 
find numerically the distribution of non-zero
matrix elements $H_{ij}\,$ . Finally, we normalize each of these
distributions to their variances and make the summation. The resulting
normalized distribution is shown in Fig.3. The envelope of this distribution
looks like the Gaussian, however, the deviations are {\it non-statistical}
ones which can be easily seen by the $\chi ^2$-test (for some bins of the
histogram, the difference is more than $100$ standard deviations).

\subsection{Density of states and spectrum statistics}

As was pointed out, the density of states $\rho (E)\,$for the two-body
random model was found to have the gaussian form \cite{FW70,BF71}. Rigorous
proof is given for the limit case of a very large number of particles and
orbitals. However, even for relatively small values of $m$ and $n$ the
distribution is very close to the Gaussian, see Fig.4. It is known that the
density of realistic physical systems such as complex atoms and heavy nuclei
can be approximated as $\rho (E)\sim \exp (A\,\sqrt{E-E_0})$ where $E_0$ is
the ground energy. Therefore, the simplest model of the two-body interaction
does not give exact correspondence to a real density, however, it
reproduces very fast increase of the density with the energy. One should
note that for full random matrices of the standard RMT, the density has the
semicircle form which is very far from the reality. It is clear that in
order to compare statistical properties of our Hamiltonian $H_{ij}\,$ with
those of complex quantum systems, one should use the left part of the energy
spectrum since the decrease of the density in the right part is due to
artificial cut-off of the single-particle spectrum (finite values of $m$ ).

\begin{figure}[htb]
\vspace{-3.0cm}
\begin{center}
\hspace{-1.5cm}
\epsfig{file=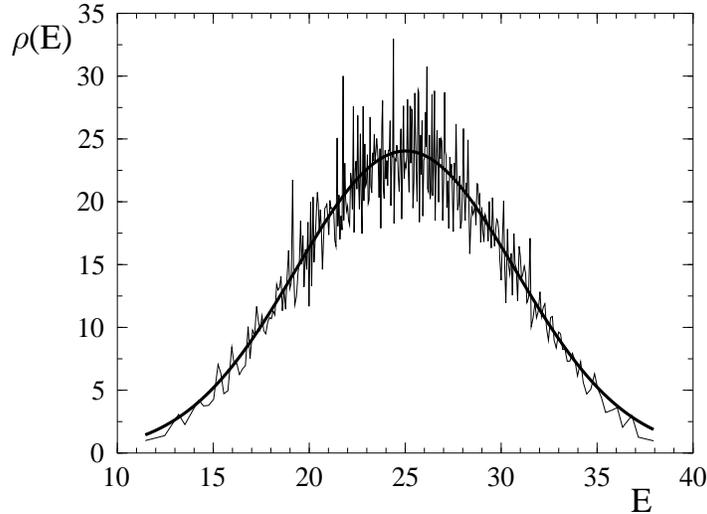,width=4.4in,height=3.6in,angle=-90}
\vspace{-1.0cm}
\caption{
Density of states of the TBRI-model with the same parameters as in
Figs.1-3. The average over $N_g=20$ is taken. The smooth curve is the best
fit to the Gaussian, with $\sigma =5.72$ and $\overline{E}=25.1$ .
}
\end{center}
\end{figure}

The interesting question is about the type of fluctuations in the energy
spectrum of the TBRI-model in comparison with those predicted by the RMT. A
particular interest is the distribution of normalized spacings $S$ between
nearest energy levels. For standard random matrices the RMT reveals specific
kind of the distribution known as the {\it Wigner-Dyson distribution}. In
particular, for symmetric real random matrices (the so-called, Gaussian
Orthogonal Ensemble of random matrices), the level spacing distribution $%
P(S) $ with a high accuracy can be described by the following form, 
\begin{equation}
\label{PS}P(S)=\frac 12\pi \,S\exp \left( -\frac{\pi S^2}4\right) 
\end{equation}
where $\left\langle S\right\rangle =1$ . In \cite{FW70,BF71} it was shown
that for the limit case of a very strong interaction, when one can neglect
the influence of the mean field (or, the same, without the leading diagonal in
the Hamiltonian $H_{ij}$ ), the form of $P(s)\,$ turns out to be quite close
to the expression (\ref{PS}). On the other hand, when studying the
distribution of spacings between the levels $E_i$ and $E_{i+k}$ with $k>1$ ,
the difference between the result of the RMT and the TBRI-ensemble is
noticeable and increases with $k$ . This means that the level spacing
distribution $P(s)\,$ is quite insensitive quantity of spectral correlations
and does not ``feel '' the difference of TBRI-matrices from the full random
matrices.

In physical applications, the interaction $V\,\,$ is typically of the same
order as the ``unperturbed '' Hamiltonian $H_0$ since in the mean field
approximation the term $H_0$ absorbs regular part of the interaction and $%
V\, $ is the (chaotic) part of the interaction which can not be included in $%
H_0$ . However, there are many cases when the interaction is weak compared
to $H_0 $ , therefore, the important question is how spectral fluctuations,
in particular, the level spacing distribution, depend on the interaction and
on 
total energy $E_i$ . The origin of the Wigner-Dyson distribution (%
\ref{PS}) is related to the onset of chaos in the exact eigenstates (see,
for example, \cite{I90}). In standard random matrices the WD-distribution
occurs for any energy since all eigenstates are completely random (their
components are distributed according to the gaussian distribution for large
size of the matrices).

Experimental data for complex atoms \cite{RP60},\cite{CG83} and heavy nuclei 
\cite{HPB82} (see also references in \cite{BG84}) agree with the
Wigner-Dyson statistics. The WD-distribution has been also observed in
numerical calculations for the Ce atom \cite{FGGK94,GFG95,FGGP99} 
and the nuclear
shell-model \cite{HZB95,FBZ96,ZBHF96}.

\begin{figure}[htb]
\vspace{-2.0cm}
\begin{center}
\hspace{-3.5cm}
\epsfig{file=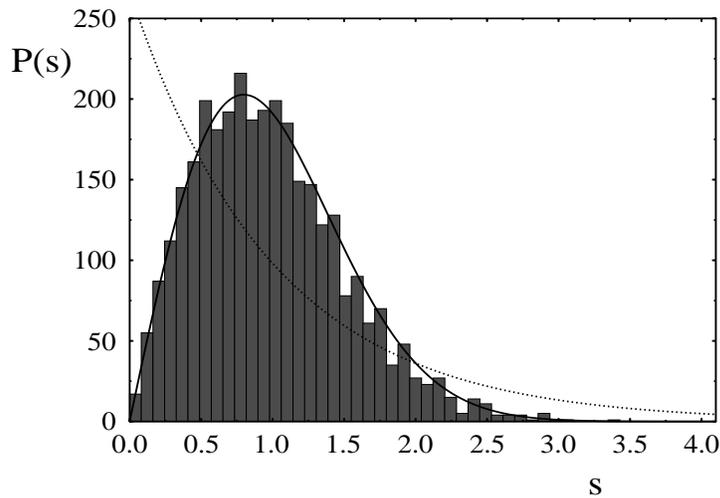,width=4.4in,height=3.6in,angle=-90}
\vspace{-2.0cm}
\caption{
 Level spacing distribution for the parameters of Figs.1-3 with $N_g=10$
 . All but $10$ levels from both edges of the energy spectrum are taken into
 account, with proper rescaling to local mean level spacings. Solid curve it
 the Wigner-Dyson distribution and the dotted curve is the Poisson, the
 latter occurs for uncorrelated energy sequences.
}
\end{center}
\end{figure}

In the TBRI-model, the randomness of the components of eigenstates is
different for different eigenstates. This fact influences the spectrum
statistics if one average over a large part on energy spectrum. In Fig.5 the
level spacing distribution for the TBRI-model is shown for the parameters
which may be compared with the Ce atom thoroughly studied in 
\cite{FGGK94,GFG95,FGGP99}, $%
n=4,\,\,m=11,\,d=0.5\,,\,V_0=0.12\,$ . The average over $N_g=10$ different
matrices $H_{ij}\,$ has been done in order to have representative
statistics. All but $10$ levels from the spectrum edges have been taken into
account. By searching the structure of eigenstates, one can see that all
eigenstates seem to be quite random (see examples below). However, one can
detect clear non-statistical (regular) deviation from the WD-distribution
which should be treated as quite strong, having in mind insensitive
character of $P(s)$ . This result may be regarded as an indication of not
strong enough interaction between particles. Detailed study of the onset of
the WD-distribution in the TBRI model has been recently performed in Ref. 
\cite{JS97}.

\subsection{Structure of exact eigenstates}

Our main interest in this Section is in the structure of exact eigenstates.
The choice of the unperturbed basis, reordered in increasing energy, allows
us to understand what happens with an increase of the interaction. In the
standard perturbation theory the natural parameter which controls the
intensity of the perturbation is the ratio $\lambda $ of the interaction to
the mean level spacing $D\,$between the unperturbed energy levels. Since the
value of $D\,$ is defined by the total density of states, $D=\rho
^{-1}\left( E\right) $ , one can expect that our control parameter is $%
\lambda =V_0/D\,$ . However, in \cite{AGKL97} it was shown that due to a
two-body character of interaction, the correct parameter is different.
Indeed, in the first order of the perturbation theory, the interaction
couples not all unperturbed states but those basis states which correspond
to the shift of not more than two particles. This means that the density of
states which are 
coupled by the two-body interaction is much less than the total
density $\rho \,\left( E\right) $ . Therefore, the correct mean level
spacing $d_f$ that has to be compared with the interaction $V_0\,$ , is
much larger than $D$ .

If the interaction is very weak, $V_0\ll d_f$ , the standard perturbation
theory can be applied. In this case any of the eigenstates in the
unperturbed basis has the form of the delta-function (originated from the
unperturbed state $\left| n_0\right\rangle $) plus small admixture of other
components with amplitudes decreasing as $\left| C_n\right| \sim 1/\left|
n-n_0\right| $ , therefore, the {\it number of principal components} is
small, $N_{pc}\sim 1$ . In this case one can speak about {\it perturbative
localization} of eigenstates in the unperturbed basis. This situation is
quite typical for eigenstates corresponding to low energies, in the energy
region where the density of states is small.

With an increase of perturbation (or when passing to higher eigenstates for
the fixed $V_0$ ), the number $N_{pc}$ of principal components with
essentially large amplitudes $C_n$ increases and can be very large, $%
N_{pc}\gg 1$ , even if $V_0\,$ is less than $d_f\,$ . Such a situation
occurs when $V_0\gg \frac 1{\pi ^2}\sqrt{D\,d_f}$ , see details in \cite
{FI97}. In such a case the structure of eigenstates is ``chaotic '',
however, there are many ``holes '' inside such eigenstates in a given basis.
Therefore, in spite of large number of components, these {\it sparse
eigenstates} are non-ergodic, which leads to {\it non-gaussian statistics}.
Namely, the fluctuations of the components $C_n\,\,$ can be extremely large
and statistical description is not valid. One should stress that in this case
the number of principal components can not be estimated as $N_{pc}\approx
\Gamma /D$ , as is typically assumed in the literature (here $\Gamma $ is an
effective ``size '' of the eigenstates in unperturbed energy
representation, see below).

When the interaction is relatively strong, $V_0\approx d_f$ , specific
transition occurs from non-ergodic to ergodic eigenstates. This transition
has been discovered in \cite{AGKL97} by considering the flow of the energy
in the Fock-space of excited states. For very large number of particles this
transition is sharp and may be compared with the Anderson transition in
solid state models (see for example details in \cite{AGKL97,MF97}). 
Therefore,
the condition $V_0>d_f$ can be considered as the transition to chaos inside
compound eigenstates, thus, allowing to describe the model in a statistical
way, see below. An example of such chaotic eigenstates is given in Fig.6 for
the parameters related to the Ce atom. One can see that for large excitation
energy ($n$ is the number of exact eigenstates reordered in increasing
energy $E^{(n)}$ ), the eigenstates look more extended (delocalized) in the
unperturbed basis. It is interesting to note that they look very similar to
the eigenstates of the Ce atom, obtained in the direct quantum computation
based on the Hartree-Fock method \cite{FGGK94,GFG95,FGGP99}. 
For the first time, chaotic
structure of eigenstates of the Ce atom has been revealed in \cite{C85}.

\begin{figure}[htb]
\vspace{-0.5cm}
\begin{center}
\hspace{-1.5cm}
\epsfig{file=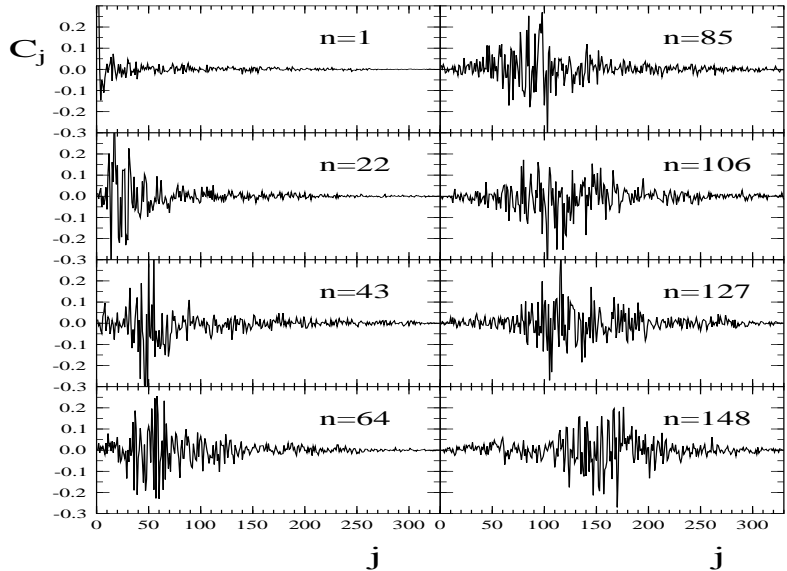,width=4.4in,height=3.6in}
\vspace{-0.0cm}
\caption{
Examples of the exact eigenstates of a matrix $H_{ij}$ for the same
parameters as in the previous figures. Components $C_j$ of $8$ eigenstates
for exact eigenstates in the low part of the spectrum are shown.
}
\end{center}
\end{figure}

Let us now discuss how to quantitatively characterize chaotic eigenstates.
First, we introduce the matrix 
\begin{equation}
\label{w}w_j^{(n)}\equiv \left| C_j^{(n)}\right| ^2=\left|
C_j(E^{(n)})\right| ^2 
\end{equation}
constructed from the exact eigenstates $\left| n\right\rangle $
corresponding to the energy $E^{(n)}$. In what follows, we use the notations
which refer low indices to the basis states, and upper indices to the exact
(compound) eigenstates. Thus, the structure of eigenstates is given by the
dependence $w_j^{(n)}$ on $j$ for fixed values of $n$ . On the other hand,
if we fix the index $j$ and explore the dependence $w_j^{(n)}\,$ on $n$ ,
one can understand how the unperturbed state $\left| j\right\rangle $ is
coupled to other basis states due to the interaction. The latter quantity is
very important since it gives the information about the spread of the
energy, initially concentrated in a specific basis state $\left|
j\right\rangle \,,$ when switching on the interaction. The envelope of this
function $w_j^{(n)}\,$in the energy representation is known as {\it strength
function }or {\it local spectral density of states }(LDOS) and will be
discussed in details in next Section.$\,$

From Fig.6 one can conclude that when the number of principal components is
large, $N_{pc}\gg 1$, such eigenstates may be treated as random
superposition of components $C_j^{(n)}$ , although they do not occupy the
whole unperturbed basis. The gaussian character of the fluctuations of $%
C_j^{(n)}$ depending on the indices $j$ or $n$ , has been revealed in \cite
{FGGK94,GFG95,FGGP99} for the Ce atom, thus allowing to treat the exact
eigenstates of a dynamical systems, as {\it chaotic eigenstates}.

The size of the basis which they occupy can be associated with the ``size ''
of eigenstates, or, with the {\it localization length}. As is known, the
notion of the localization length is very important in solid state
applications, when studying the eigenstates of disordered models in infinite
basis in the position representation. In such applications the localization
length $l_\infty \,$ is defined via exponential decrease of the square of
the amplitude of eigenstates . For the finite basis, the definition of the
localization length can be generalized in a way described in \cite{I90}.
Following to \cite{I90,FM94} we define here two localization lengths,
namely, the ``{\it entropy localization length}'' $l_h$, and the
localization length $l_{ipr}\,$associated with the so-called {\it %
participation ratio}. The first one is defined by the expression 
\begin{equation}
\label{lh}l_h=N\exp ({\cal \left\langle H\right\rangle }-{\cal H}_0), 
\end{equation}
where ${\cal \left\langle H\right\rangle }$ is the mean `` {\it entropy}''
of eigenstates, 
\begin{equation}
\label{entropy}{\cal \left\langle H\right\rangle }=-\frac
1M\sum_{n=1}^M\sum_{j=1}^Nw_j^{(n)}\ln \,(w_j^{(n)})\, 
\end{equation}
and ${\cal H}_0$ is the normalization constant which is equal approximately
to $2.08$ in the case of pure gaussian fluctuations of $C_j$ (see details
e.g. in \cite{I90}). Here $M$ is the number of eigenstates which are taken
for the average. This can be the number of eigenstates of one matrix $H_{ij}$
taken from a small energy window, or the number of eigenstates for the fixed 
$n$ , computed from different matrices $H_{ij}\,$ with different realization
of disorder in two-body matrix elements.

The second definition is commonly used in solid state applications (see e.g. 
\cite{FM94}). Assuming the gaussian character of fluctuations of the
components of eigenstates, the localization length $l_{ipr}$ is defined by 
\begin{equation}
\label{lipr}l_{ipr}=\frac 3P\,;\,\,\,\,\,\,\,\,\,\,P=\frac
1M\sum_{n=1}^M\sum_{j=1}^N\,\left( w_j^{(n)}\right) ^2 
\end{equation}
In the above expressions (\ref{lh}, \ref{lipr}) the factors $2.08$ and $3$
are, in fact, normalizing coefficients which provide, in the limit case of
completely extended and gaussian eigenstates in the finite basis of the size 
$N$ , the ``maximal'' value of the localization length $l_h=l_{ipr}=N$. In
other extreme case of a strong (exponential) localization in the unperturbed
basis, the above two localization lengths are proportional to that found
from the tails of eigenstates (see \cite{I90}). The dependence
of the localization lengths $l_h$ and $l_{ipr}$ on the basis number and
energy of exact eigenstates is given in Fig.7a-b. The average over $N_g=50$
matrices $H_{ij}$ with different realizations of two-body matrix elements
has been taken, in order to smooth strong fluctuations in the value of
localization lengths of individual eigenstates. The above definitions of
localization length can be taken for the estimate of a number of principal
components $N_{pc}$ , and can be associated with the degree of ``chaoticity
'' of compound eigenstates. The data of Fig.7 show quite good correspondence
with direct computations \cite{FGGK94,GFG95} 
performed for the Ce atom, for which
the localization length $l_h$ was found to be about $l_h\approx 110-130$.
Comparing Fig.7b with Fig.4 of the density of states $\rho \,(E)$ , one can
see the similarity. This fact can be understood from the simplest estimate
of $N_{pc}$ as a total number of (basis) states defined by the ``width'' $%
\Gamma $ , the latter can be approximated as the mean-square-root of the
distribution $w_j^{(n)}$ for the fixed $n$ , 
\begin{equation}
\label{NpcG}N_{pc}\approx \frac \Gamma D=\Gamma \rho \,(E) 
\end{equation}
One should stress that this expression is valid for ergodic eigenstates ( $%
V_0\geq d_f\,$ ) and shows the proportionality of $N_{pc}$ to the density.
As for the width $\Gamma $ , it is approximately independent of the
excitation energy, see below.

\begin{figure}[htb]
\vspace{-1.0cm}
\begin{center}
\hspace{1.5cm}
\epsfig{file=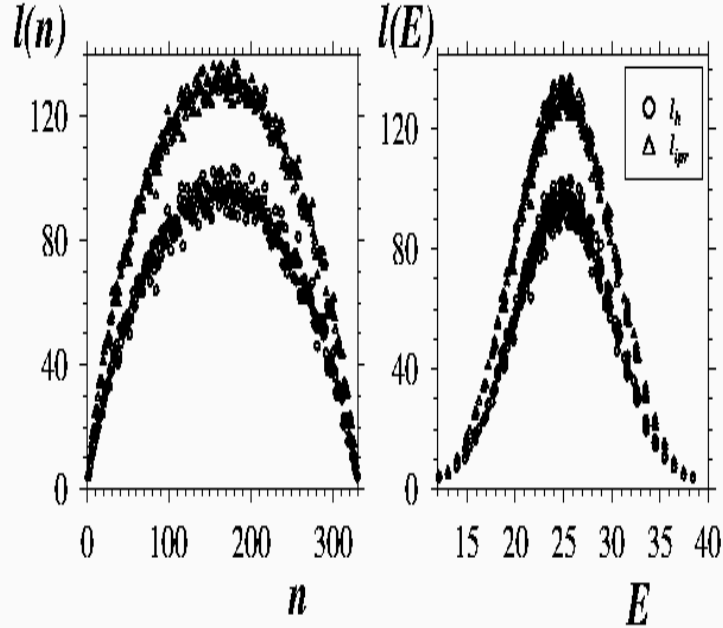,width=4.in,height=4.in,angle=-90}
\vspace{0.0cm}
\caption{
(a) Localization lengths $l_{h}$ and $l_{ipr}$ are given in
dependence on the number $n$ of exact eigenstates $\left| n\right\rangle $ ,
see (\ref{lh}) and (\ref{lipr}). The parameters are taken the same as in
Figs.1-6, with $N_g=50$ . (b) the same as in (a), but in the energy
representation.
}
\end{center}
\end{figure}

Eq.(\ref{entropy}) reflects general relation between the number of principal
components $N_{pc}$ and the {\it entropy of eigenstates}, 
\begin{equation}
\label{SEF}S_{EF}=\ln \,N_{pc} 
\end{equation}
where $S_{EF}\,$ stands for any reasonable definition of the entropy. In
application to shell models of complex nuclei this relation has been studied
in great details in \cite{HZB95,ZBHF96}. Combining Eq.(\ref{SEF}) with
Eq.(\ref{NpcG}), one can get 
\begin{equation}
\label{SEFG}S_{EF}\approx \ln \,\rho \,(E)\,+\,\ln \,\Gamma (E)\, 
\end{equation}
In contrast to the density of states, the width $\Gamma \,$ is a weakly
dependent function of the energy. Therefore, the entropy $S_{EF}$ found from
exact eigenstates practically coincides with the thermodynamical entropy,
the fact which was mentioned for the first time in \cite{HZB95,ZBHF96}.

\subsection{Strength function}

In this Section we discuss the properties of \thinspace the {\it strength
function }which is defined as 
\begin{equation}
\label{LDOS}W(E^{(m)},j)=\sum_n|C_j^{(n)}|^2\delta (E-E^{(n)}) 
\end{equation}
Here $C_j^{(n)}$ are the components of eigenstates $\left| n\right\rangle $
(''compound'' states) of the total Hamiltonian $H$ given in the unperturbed
basis $\left| j\right\rangle $ , and $E^{(n)}$ is the energy associated with
the state $\left| n\right\rangle $. The sum is taken over a number of
eigenstates $\left| n\right\rangle \,$ chosen from a small energy window
centered at the energy $E\,^{(m)}$ . One can see that this function $%
W(E,j)\, $ is originated from the same matrix $w_j^{(n)}$ which has been
introduced in previous Section when discussing the structure of exact
eigenstates. Indeed, an exact eigenstate is characterized by the dependence $%
w_j^{(n)}$ on $j$ for the fixed value $n$ (associated with the energy $%
E^{(n)}$ ). On the other hand, the strength function is characterized by the
same function $w_j^{(n)}$ when index $j$ is fixed and we are interested in
the dependence on the energy $E^{(m)}$ due to the relation between $%
E^{(n)}\, $ and $n$ therefore, 
\begin{equation}
\label{Fdef}W(E^{(m)},j)\simeq F_j^{(n)}\rho
(E),\,\,\,\,\,\,\,\,\,\,\,\,\,F_j^{(n)}\equiv \overline{w_j^{(n)}}. 
\end{equation}
Here we have introduced the $F-${\it function} $F_j^{(n)}$ which gives the
envelope of $w_j^{(n)}$ in dependence on the indices $j$ and $n$ (the
bar stands for the average inside small windows centered at $j$ and $n$). In
fact, the strength function is the (smooth) representation of a (simple)
basis state $\left| j\right\rangle $ in terms of exact eigenstates. This
function is very important since it can be measured experimentally. It
contains an information about the internal interaction between unperturbed
states. Namely, it shows how the unperturbed state $\left| j\right\rangle $
is coupled to the exact states $\left| n\right\rangle $ due to the
interaction. An effective width of this function ({\it spreading width })
defines the energy range associated with the ``life time'' of an unperturbed
state $\left| n\right\rangle $ if initially one excites specific basis state.

In solid state models the role of the unperturbed energy in Eq.(\ref{LDOS})
plays the position $j$ of an electron and the function $W(E^{(m)},j)%
\Rightarrow W(E,j)\,\,$ has the meaning of the electron density of states
for the fixed position $j$ . For this reason this function is known in solid
state physics as the {\it local density of states }(LDOS). One can see that
if (apart from fluctuations) $w_j^{(n)}$ is independent of the position (or,
in our application, the energy of compound state), it reduces to the total
density of states, $W(E^{(m)},j)\Rightarrow \rho \,(E)$ .\thinspace Also, if
the total density of states is constant, $\rho =$ $\alpha ^{-1}$ , the
dependence of the LDOS on $j$ is of the form $W(E,n)=W(E-\alpha j)$ . This
means that the form of the LDOS is the same for any $j$ . In our
case of strong dependence of the density of states on the energy, the form
of the LDOS is quite complicated, see next Section. Normalized to the mean
energy level spacing, the strength function $W(E^{(m)},j)$ determines an
effective number $N_{pc}$ of principal components of compound states $\left|
n\right\rangle $ which are present in the basis state $\left| j\right\rangle 
$.

Similar to the analysis of the structure of the eigenstates in dependence on
the interaction, one can understand that for a very weak interaction $V_0\ll
d_f$ the LDOS\ is a delta-like function with a very small admixture of other
components which can be found by the standard perturbation theory. With an
increase of the interaction, the number of principal components increases
and can be very large. However, if the interaction is not strong enough \cite
{FI97}\thinspace , 
\begin{equation}
\label{V0d}1\gg \frac{V_0}{d_f}\gg \frac 1{\pi ^2}\sqrt{\frac D{d_f}}\,\,\,, 
\end{equation}
the LDOS\ is sparsed, with extremely large fluctuations of components, see
details in Ref.\cite{MF97}. In order to have ergodic LDOS, one needs to have
the perturbation large enough, $V_0\gg d_f\,$(for a large number of
particles this transition is sharp and, in fact, one needs the weaker
condition, $V_0\geq d_f$ \thinspace , see details in \cite{FI97}).

\begin{figure}[htb]
\vspace{-0.7cm}
\begin{center}
\hspace{-2.3cm}
\epsfig{file=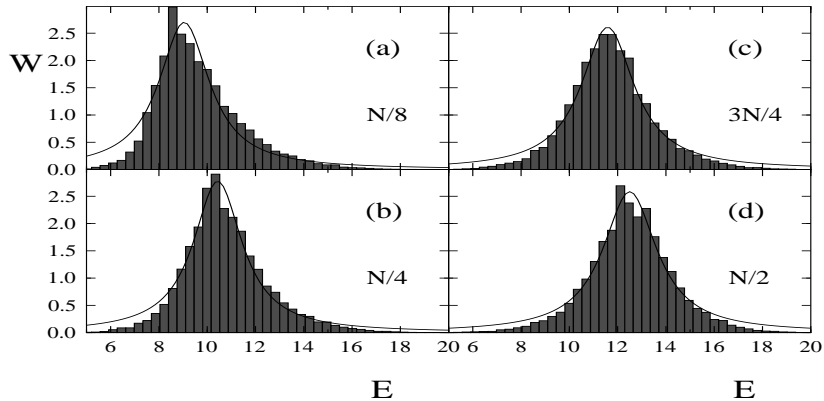,width=4.4in,height=3.6in}
\vspace{-2.5cm}
\caption{
Few examples of the LDOS for the TBRI-model for different basis states
$j=N/8,\,N/4,\,3N/4,N/2$ . The average is done over a number of eigenstates
with close energies, for $n=4\,$ particles, $m=11$ orbitals, $%
V_0=0.12,d_0=0.5$ and with additional average over $N_g=50$ matrices $H_{ij
}$ . Smooth curves are the best fit to the Gaussian at the center of
the energy spectrum, $j=N/2$ .
}
\end{center}
\end{figure}

For $V_0\gg d_f$ the LDOS\ turns out to be ergodic 
and thransition to chaos occurs \cite
{AGKL97,SS97,MF97,WPI97,JS97,FG94,S97,S98}, therefore, 
statistical description of
the model is valid. Few examples of the LDOS for the TBRI-model are given in
Fig.8 for the parameters of the Ce atom. One can see strong dependence of
the shape of the LDOS\ on the position of basis state $\left|
j\right\rangle $ in the energy spectrum.

Before we start with the discussion of analytical results, it is important
to point out the correspondence between the shape of the LDOS, and that of
exact eigenstates which have been discussed in previous Section.
Specifically, from the analysis of the structure of the eigenstate matrix $%
w_j^{(n)}\,\,\,$ one can expect the similarity between the LDOS\ and shape
of the eigenstates. For the first time such a similarity has been observed
when studying band random matrix ensembles \cite{CCGI93,CCGI96}. Moreover,
the detailed study of some dynamical models \cite{BGI98,WIC98} have revealed
that even when the shapes of the LDOS and EFs seem to be completely
different, after a proper rescaling which involves both unperturbed and
perturbed energy spectrum, both shapes are very similar. In order to compare
characteristics of the LDOS\ and EFs, in Fig.9 the dependence of the entropy
localization length on the energy is shown for both the LDOS\ and EFs. Apart
from strong fluctuations (which are due to a chaotic nature of the
components $C_j^{(n)}$ ), in general, the dependencies $l_H\left( E\right)
\, $ look very similar.

The problem of the correspondence between shapes of the LDOS and EFs is
still open, however, from the studies made up to now (see also \cite
{BGI98,WIC98}), one can conclude that if the width of the perturbed spectrum
is of the same order as the unperturbed one (or, the same, the perturbation
is not very larger), one can expect that both shapes are very close to each
other. The importance of this problem of similarity for the shapes of the
LDOS and EFs will be clear in next Section when we discuss the relation
between the shape of the EF and generic properties of the occupation number
distribution.

\begin{figure}[htb]
\vspace{-2.0cm}
\begin{center}
\hspace{-1.5cm}
\epsfig{file=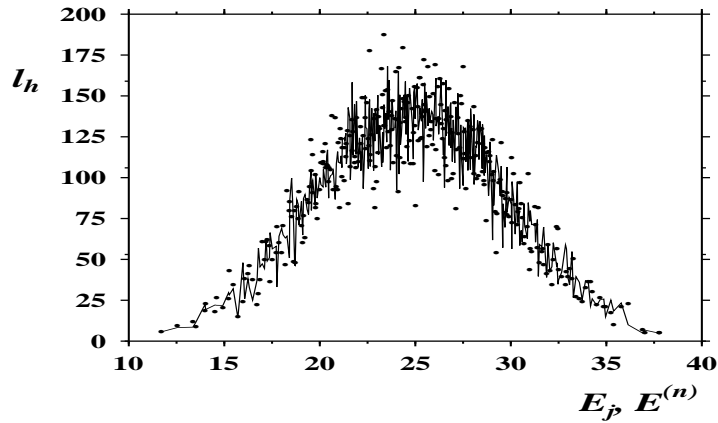,width=3.0in,height=4.3in,angle=-90}
\vspace{0.2cm}
\caption{
Comparison of the localization length $l_h$ for the LDOS (points) and
eigenstates (line). The data are given for one Hamiltonian matrix with the
same parameters as in Figs.1-7, for the same numbers $j=n$ .
}
\end{center}
\end{figure}

For the first time the form of the LDOS in random matrix theory has been
discussed in Ref. \cite{W55} where band random matrix ensemble has been
considered. The matrices were assumed to consist of the diagonal part $H_0$
in the form of reordered numbers $H_{jj}=j\,D$ (thus, the unperturbed level
density is constant, $\rho _0=D^{-1}$ ), and the perturbation $V_0$ with the
random independent off-diagonal matrix elements for $\left| i-j\right| \leq
b $. Outside the band of size $b\gg 1$ matrix
elements are zeros. The distribution of the off-diagonal matrix elements is
characterized by the zero mean, $\left\langle H_{ij}\right\rangle =0\,$ and
the variance $V_0^2=\left\langle H_{ij}^2\right\rangle \,.$ For this model,
the relevant parameter was found to be

\begin{equation}
\label{q}q=\frac{\rho _0^2V_0^2}b. 
\end{equation}
Wigner analytically proved \cite{W55} that for relatively strong
perturbation, $V_0\gg D\,\,$ in the limit $q\ll 1\,$ (when the influence of
the main diagonal $H_{jj}$ is strong) the form of the LDOS is the Lorentzian,

\begin{equation}
\label{BW}W_{BW}\,(\tilde E)=\frac 1{2\pi }\,\frac{\Gamma _{BW}}{\tilde E^2+ 
\frac{\Gamma _{BW}^2}4},\,\,\,\,\,\,\,\,\tilde E=E-D\,n 
\end{equation}
which is nowadays known as the {\it Breit-Wigner} (BW) form. Here $\Gamma
_{BW}$ is the spreading width which is half-width of the distribution (%
\ref{BW}), 
\begin{equation}
\label{BWgam}\Gamma _{BW}=2\pi \rho _0V_0^2 
\end{equation}
and the energy $\tilde E=E-D\,n$ refers to the center of the distribution.
In other limit $q\gg 1\,$ the influence of the unperturbed part $H_0\,$ can
be neglected and the shape of the LDOS\ tends to the shape of the total
density of states of band random matrices without the leading diagonal $%
H_{jj}$, which is the semicircle.

Recently, the shape of the LDOS has been studied rigorously for a more
general distribution of the off-diagonal elements $v_{nm}$ in the Wigner band
random matrix model (WBRM). Namely, the variance of random matrix elements
is taken to depend on the distance $r=\left| n-m\right| $ from the principal
diagonal according to an envelope function $f(r)$ which, for $r\rightarrow
\infty $, decreases sufficiently fast (see details in \cite{FCIC96}). In
this case the effective band size $b\,$ is defined by the second moment of
the function $f(r)$ . Another important generalization of Ref.\cite{FCIC96}
is related to the sparsity inside the band, which can be defined as a
relative number of zero elements in each line of the matrix. As was shown
above, such a sparsity, which is due to the two-body nature of the
interaction between the particles, is important for statistical properties
of compound states.

The BW-form of the LDOS also occurs in a more general model describing the
interaction of an unperturbed state with a large set of complex states, see
details in the book \cite{BM69}. It also appears for dynamical systems with
complex behavior, such as the Ce atom \cite{FGGK94,GFG95,FGGP99}, 
the $sd$ shell model 
\cite{HZB95,FBZ96,ZBHF96}, as well as in random models in
application to solid state physics \cite{GS97,MRRW98}.

For a long time it was believed that the LDOS for complex physical systems
has the universal energy dependence described by the BW-distribution (\ref
{BW}). However, when studying the structure of the LDOS\ and EFs of the Ce
atom, it was observed some deviation from the BW-shape for large distance
from the center. By applying the WBRM, in \cite{FGGK94} it was analytically
shown that the shape of the LDOS is highly non-universal for energies larger
than the effective size of the band in the energy representation, $\left|
E\right| \gg D\,b$ (in what follows by $E\,$ we mean the distance from the
center $E_c$ of the distribution, $\tilde E\Rightarrow E$ ), see also
numerical data in \cite{FCIC96}. Namely, outside this range, the tails of
the LDOS are highly non-trivial, decaying very fast (even faster than the
exponent) when $\left| E\right| \rightarrow \infty $ . Therefore, the range
of parameters for which the form of LDOS has the BW-form in the WBRM-model,
is given by the condition \cite{FCIC96} which can be written as

\begin{equation}
\label{range2}\rho _0^{-1}\,\ll \,\Gamma _{BW}\ll \,b\rho _0^{-1} 
\end{equation}
Here, the left-hand-side of the inequality is related to non-perturbative
character of the coupling since the perturbation should couple many
unperturbed states.

The condition (\ref{range2}) turns out to be of generic and can be applied
to real physical systems. Moreover, it can be used to find the effective
band-width $b_{eff}$ of interaction when other definitions of $b\,$ are
obscure. In the TBRI-model with many interacting particles the band-width is
very large (practically it is infinite), and does not play
any role in many solid state applications. On the contrary, in application
to complex atoms and nuclei, the number of particles above the Fermi level
is relatively small (four particles in the Ce atom and 12 particles in the $%
sd$ shell models), and this effect can be important. It should be pointed
out that the BW-dependence has infinite second moment. At the same time, in
any physical application the second moment is always finite and is defined by
the sum of the square of the off-diagonal elements. This fact results in
non-Lorentzian tails of the LDOS.

The important question is about the shape of the LDOS for a strong
interaction, when the condition (\ref{range2}) violates, $\Gamma
_{BW}\approx \,b\rho _0^{-1}$ . Since in the mean field approximation the
``regular '' part of the interaction is included in the mean field, very
often the interaction $V$ is compared with the unperturbed part $H_0$ .
Therefore, this situation when the shape of the LDOS\ is very different from
the BW-form, is quite physical. In order to understand what happens in this
case, it is useful to study how the transition from the BW-form to the
semicircle occurs in the WBRM-model. As was pointed out, the semicircle form
itself is unphysical and appears when neglecting the unperturbed part $H_0$
. However, the form of the LDOS in the transition region seems to be of
quite generic.

Numerical data in Refs.\cite{FCIC96},\cite{CFI99} for the WBRM-model have
shown that for the case when $\Gamma _{BW}\approx \,b\rho _0^{-1}$ the form
of the LDOS can be approximately described by the Gaussian. The same fact
has been observed and discussed in Refs.\cite{HZB95,ZBHF96} when
studying eigenstates and LDOS for shell model of nuclei. Since in the latter
application the band size of the interaction is not well defined, it is
better to introduce one more parameter, in addition to the half-width $%
\Gamma _{BW}\,\,\,$ . As was shown in Ref.\cite{CFI99}, in general, the form
of the LDOS can be effectively described by two independent parameters, $%
\Gamma _{BW}$ and $\Delta E\,$ where the latter is defined via the variance
of the LDOS, 
\begin{equation}
\label{delEint}\left( \Delta E\right) ^2=4\sigma _W^2=4\int \left(
E-E_c\right) ^2W_{BW}\,(E)\,dE 
\end{equation}
In the symmetric case (in the TBRI-model, at the center of the
spectrum), $E_c$ coincides with the unperturbed energy, $E=E_j$ .

Then, if the parameter $\Gamma _{BW}$ is much less than 
$\Delta E$ , the shape $W\,(E)$ of the LDOS is the
BW-dependence (\ref{BW}) and $\Gamma _{BW}$ has the meaning of the
half-width of the distribution $W(E)=W_{BW}\,(E)$ . On the other hand, if $%
\Gamma _{BW}\approx \Delta E$ , the form of the LDOS is approximately the
Gaussian, 
\begin{equation}
\label{WGauss}W(E)=\frac 1{\sigma _W\sqrt{2\pi }}\exp \left( -\frac{\left(
E-E_c\right) ^2}{2\left( \sigma _W\right) ^2}\right) 
\end{equation}
Detailed numerical study \cite{CFI99} of the form of the LDOS in the region $%
\Gamma _{BW}\approx \Delta E$ have shown that the LDOS\ coincides with the
Gaussian with a very high accuracy. It is important to stress that in this
case, the parameter $\Gamma _{BW}$ has nothing to do with the half-width $%
\Gamma _{hw}$ of the LDOS, the latter is proportional to the
mean-square-root $\sigma _W$ of the Gaussian, $\Gamma _{hw}\approx
C_0\,\sigma _W\,$ (see also discussion in \cite{HZB95,ZBHF96}). Note
that the Gaussian form typically occurs in ``statistical spectroscopy'' \cite
{FW70,DFW77,FKPT88} 
when neglecting the mean field term $H_0$ in Eq.(\ref{H}). One should
note that sometimes the fact that the LDOS can deviate from the BW-form due
to the influence of the finiteness of $\Delta E\,$ , is missing in the
literature.

It is important that the center and variance of the LDOS can be explicitly
expressed via diagonal and off-diagonal matrix elements respectively \cite
{FI97}. Indeed, the center is defined by%
$$
E_c=\left( \overline{E^{(n)}}\right) _j=\sum_nE^{(n)}F_j^{(n)}\approx
\sum_nE^{(n)}w_j^{(n)}= 
$$
\begin{equation}
\label{Ecent}\sum_{n,m}\left\langle j\right| \left. n\right\rangle
\left\langle n\right| H\left| m\right\rangle \,\left\langle m\right| \left.
j\right\rangle =H_{jj}=E_j 
\end{equation}
where the relation $\left\langle i\right| H\left| j\right\rangle =\delta
_{ij}\,\left\langle i\right| H\left| j\right\rangle $ is used for exact
eigenstates. Correspondingly, the variance can be obtained from the matrix
elements of $H^2$ , 
\begin{equation}
\label{sigmdef}\left( \sigma _W^2\right) _j=\sum_n\left( E^{(n)}-E_j\right)
^2F_j^{(n)}\approx \sum_n\left( E^{(n)}-E_j\right) ^2w_j^{(n)}=\sum_{p\neq
j}H_{jp}^2 
\end{equation}
For the TBRI-model the sum of the off-diagonal elements for any fixed value
of $p$ can be evaluated exactly \cite{FGI96}, 
\begin{equation}
\label{H2}\frac{\left( \Delta E\right) _j^2}4=\sum_{p\neq
j}H_{jp}^2=V_0^{2\,}\left( n-1\right) K_1+V_0^2\,K_2=\frac
14V_0^{2\,}n\left( n-1\right) \left( m-n\right) \left( 3+m-n\right) 
\end{equation}
where the expressions (\ref{K123}) have been used for $K_1\,$and $K_2$ , and 
$V_0^2$ is the variance of the off-diagonal matrix elements. One can see
that with an increase of the interaction, the half-width of the LDOS changes
from the quadratic dependence $\Gamma _{hw}\sim V_0^2$ to the linear one, $%
\Gamma _{hw}\sim V_0$ . It is interesting to note that the variance of the
LDOS for Fermi-particles does not depend on the index $j$ which stands for a
specific basis state, therefore, $\left( \Delta E\right) _j^2\Rightarrow
\left( \Delta E\right) ^2$

\subsection{Analytical solution for the LDOS}

Very recently, the form of the LDOS for the TBRI-model has been analytically
found in \cite{FI99} for any strength of perturbation. In this Section we
discuss the approach of \cite{FI99} and the obtained results. We would like
to stress that our aim is to find the LDOS in terms of matrix elements of
the total Hamiltonian $H_{ij}$ , without its diagonalization. To start with,
let us rewrite the general expression for the LDOS in the form

\begin{equation}
\label{Ff}W_k\,(E)=F(E_k,E)\,\rho (E) 
\end{equation}
where $E$ is the total energy of the system (energy of an exact eigenstate).
As was pointed out, the $F-$function gives the shapes of both exact
eigenstates and strength functions depending on what is fixed, the total
energy $E\equiv E^{(i)}$ or the unperturbed one, $E_k$. The method used in 
\cite{FI99} is an extension of the approach developed in \cite{BM69,LBBZ95},
which takes into account specific structure of the Hamiltonian TBRI-matrix.
Specifically, first, we fix some basis component $\left| k\right\rangle \,$
and diagonalize the Hamiltonian matrix without this component. Then the
problem is reduced to the interaction of this component with exact
eigenstates $\left| i\right\rangle \,$ which are statistically described by
the matrix components $V_{ki}$. In the spirit of the approach of Ref.\cite
{BM69} let us introduce a small energy window $\Delta $ which will be used
for an average over the total energy inside this interval. As a result, the
set of equations for the LDOS can be written in the following form, 
\begin{equation}
\label{FfBW}W_k(E)=\frac 1{2\pi }\frac{\Gamma _k(E)}{(E_k+\delta
_k-E)^2+\frac 14\,\Gamma _k^2(E)} 
\end{equation}
where 
\begin{equation}
\label{Gamma}\Gamma _k(E)\simeq 2\pi \overline{\left| V_{ki}\right| ^2}\rho
(E) 
\end{equation}
is some function which can be associated with the half-width of the
distribution $W_k(E)\,$ , for the case when the energy dependence is weak, $%
\Gamma _k(E)\simeq const\,$ . The energy shift $\delta _k\,$ for the basis
state $\left| k\right\rangle $ , 
\begin{equation}
\label{delta}\delta _k=\sum_i\frac{\left| V_{ki}\right| ^2(E-E^{(i)})}{%
(E-E^{(i)})^2+\frac{\Delta ^2}4} 
\end{equation}
is due to the asymmetry of the perturbation, if the energy $E_k\,$is not at
the center of the spectrum. This shift is just the modified second order
correction to the unperturbed energy level. For the calculation of the shape
of the eigenvector $|i>$ one should substitute the exact energy $%
E=E^{(i)}=E_i+\delta _i$. Then, if the interaction is not very strong, in
the evaluation of the above equations the difference $\delta _i-\delta _k$
can be neglected.

One should stress that the summation in the above equations is performed
over exact states. Since the exact eigenstates are unknown, one should
express everything in terms of the basis states only. To do this, we express
exact eigenstates $|i>$ through the basis components, 
\begin{equation}
\label{Vki}|V_{ki}|^2=\sum\limits_p\left| C_p^{(i)}\right|
^2|H_{kp}|^2+\sum\limits_{p\neq q}C_q^{(i)*}C_p^{(i)}H_{kp}H_{qk} 
\end{equation}
In previous Sections it was argued that for large number of principal
components, $N_{pc}\gg 1\,$and sufficiently strong interaction $V_0\geq d_f$
, the components $C_{p,q}^{(i)}$ can be treated as random variables,
therefore, the second term in (\ref{Vki}) vanishes after averaging.
Substitution of Eq. (\ref{Vki}) into Eqs.(\ref{Gamma}, \ref{delta}) gives 
\begin{equation}
\label{Gammaf}\Gamma _k(E)=2\pi \sum\limits_{p\neq
k}|H_{kp}|^2W_p(E)=\sum\limits_{p\neq k}|H_{kp}|^2\frac{\Gamma _p(E)}{%
(E_p+\delta _p-E)^2+\frac{\Gamma _p^2(E)}4} 
\end{equation}
\begin{equation}
\label{deltaf}\delta _k=\sum\limits_{p\neq k}\left| H_{kp}\right| ^2\int
dE^{(i)}\frac{W_p(E^{(i)})}{E-E^{(i)}}\simeq \sum\limits_{p\neq k}\frac{%
\left| H_{kp}\right| ^2(E-E_p-\delta _p)}{(E-E_p-\delta _p)^2+\frac{\Gamma
_p^2(E)}4} 
\end{equation}
where the integral is taken as the principal value. Last equality is valid
in the approximation of slow variation of $\Gamma _p(E)$ and $\delta _p$.
The equations for $\Gamma _k(E)$ and $\delta _k$ allow to calculate the
strength function (\ref{FfBW}) from the unperturbed energy spectrum and
matrix elements of the total Hamiltonian $H$.

Now we have the set of equations (\ref{FfBW},\ref{Gammaf},\ref{deltaf})
which, in principal, give the solution for the LDOS $W_k(E)$ and can be
solved numerically. However, for relatively large number of particles
(practically, for $n\ge 4$), one can find an approximate analytical solution
of the problem \cite{FI99}. By analyzing these equations, in \cite{FI99} it
was proved that the condition of the self-consistent solution for the LDOS
(or the same, when the shape of the LDOS exists as a smooth function of the
energy), is just the condition for the onset of chaos in the TBRI-model, $%
V_0\geq d_f$ , discussed in previous Sections.

More specifically, for a very strong interaction, $\Gamma >>d_f$ the number $%
N_f$ of effectively large terms in the sums is large, $N_f\sim \Gamma /d_f$,
fluctuations of $\Gamma $ are small, $\delta \Gamma \sim \Gamma /\sqrt{N_f}$
and Eq. (\ref{Gammaf}) can be written in the form, 
\begin{equation}
\label{GammaH}\Gamma _k(E)\simeq 2\pi \overline{\left| H_{kp}\right| ^2}\rho
_f\,(\tilde E) 
\end{equation}
Here $\tilde E=E-\delta $ and the energy shift $\delta \equiv \,<\delta _p>$
can be neglected in the case of $\Gamma <<\sigma $ with $\sigma $ standing
for an effective band-width $\sigma $ of the Hamiltonian matrix $H_{pq}$
(see Eq.(\ref{rhof})). In order to perform the summation over $p$, it was
assumed that $\Gamma (E)$ and $\rho _f\,(E)$ change slowly within the energy
interval of the size $\Gamma $. As a result, in order to have large number
of final states $N_f\sim 2\pi H_{kp}^2/d_f^2$ and {\it statistical
equilibrium} (small fluctuations of $\Gamma $), one needs $H_{kp}>>d_f$. In
this case chaotic components of exact eigenfunctions in the unperturbed
many-particle basis ergodically fill the whole energy shell of the width $%
\Gamma $, with Gaussian fluctuations of the coefficients $C_k^{(i)}$ and the
variance given by the $F-$function (\ref{Fdef}) (see also 
\cite{BM69,FGGK94,FGGP99}%
).

With a decrease of the ratio $H_{kp}/d_f$ , the fluctuations of $\Gamma $
increase and for $H_{kp}<d_f$ the smooth self-consistent solution of Eqs.(%
\ref{Gammaf}) does not exist. Indeed, in this case the term $\Gamma _p$ in
the denominator of Eq.(\ref{Gammaf},\ref{deltaf}) can be neglected and the
sum in (\ref{Gammaf}) is dominated by one term with the minimal energy $%
E-E_p\sim d_f$. Therefore, for a typical basis state $|k>$ formally one gets 
$\Gamma _k\sim \Gamma _p(H_{kp}/d_f)^2<<\Gamma _p$. This contradicts to the
assumption of the equilibrium according to which all components are of the
same order, $\Gamma _k\sim \Gamma _p$.

One should stress again that the absence of a smooth solution for the shape
of the eigenstates and the strength function does not mean that the number
of principal components $N_{pc}$ in exact eigenstates is small. It can be
large, however, the distribution of the components is not ergodic, there are
many ``holes'' inside exact eigenstates which occupy the energy shell of the
width $2\pi \overline{\left| H_{kp}\right| ^2}\rho _f(E)$ (see \cite
{AGKL97,FI97}). In such a situation, the fluctuations of $C_k^{(,i)}$ are
very large and non-Gaussian.

It is important to note that the ensemble average (over many matrices $%
H_{kp} $ ) in this problem is not equivalent to the energy average (inside
specific Hamiltonian matrix). Indeed, the average over the single-particle
spectrum leads to the variation of energy denominators in (\ref{Gammaf}) and
can fill the ``holes '' in the $F-$function.

From general equations for the shape of the LDOS one can make an unexpected
conclusion that the spreading width $\Gamma (E)$ can be a strong function of
excitation energy $E$ due to the variation of the density $\rho
_f\,(E)=d_f^{-1}$ of final states in Eq. (\ref{GammaH}). In Ref.\cite{FI99}
it was shown that for the excited states, well above the ground state, the
energy dependence of $\rho _f\,(E)$ and $\Gamma (E)$ can be quite close to
the Gaussian. This result is based on the estimate of the two-body density $%
\rho _f\,(E)$ , 
\begin{equation}
\label{rhosum}\rho _f\,(E)=\rho _f^{(1)}(E)+\rho _f^{(2)}(E) 
\end{equation}
where the density $\rho _f$ is determined by the energy difference $\omega
_{pk}^{(2)}$ between the states $|p>$ and $|k>$ which differ by the position
of two particles, and by $\omega _{pk}^{(1)}$ between those states which
differ by the position of one particle. Detailed analysis \cite{FI99} have
shown that both $\rho _f^{(1)}(E)$ and $\rho _f^{(2)}(E)$ for large number
of particles are described by the Gaussian, 
\begin{equation}
\label{rhof}\rho _f^{(1,2)}(\tilde E)\simeq \frac{K_{1,2}}{\sigma _{1,2} 
\sqrt{2\pi }}\,exp\left( -\frac{\left( \tilde E-E_k-\overline{\omega ^{(1,2)}%
}\right) ^2}{2\sigma _{1,2}^2}\right) 
\end{equation}
Normalization parameter $K_{1,2}$ stands for the number of one or
two-particle transitions, see Eq.(\ref{K123}). Here the average frequency of
one and two-particle transitions reads as 
\begin{equation}
\label{omega1}\overline{\omega ^{(1)}}\approx m/(m-n)(\overline{\epsilon }%
-E_k/n) 
\end{equation}
and 
\begin{equation}
\label{omega2}\overline{\omega ^{(2)}}=2(\overline{\epsilon _p}-\overline{%
\epsilon _k})\approx 2m/(m-n)(\overline{\epsilon }-E_k/n) 
\end{equation}
where $\overline{\epsilon _k}=E_k/n$ is the mean single-particle energy in
the basis state $|k>$ containing $n$ particles, $\overline{\epsilon }$ is
the single-particle energy averaged over all $m$ orbitals, and the mean
energy of the empty orbitals $\overline{\epsilon _p}$ can be found from the
relation $m\overline{\epsilon }=\overline{\epsilon _k}n+\overline{\epsilon _p%
}(m-n)$.

The variance of $\rho _f^{(1,2)}(E)$ for one and two-particle transitions is 
\begin{equation}
\label{sigma1}\sigma _1^2=\sigma _p^2+\sigma _k^2+2(n-1)V^2\approx \sigma
_\epsilon ^2+2(n-1)V^2 
\end{equation}
and 
\begin{equation}
\label{sigma2}\sigma _2^2=2\sigma _p^2+2\sigma _k^2+(4n-6)V_0^2\approx
2\sigma _\epsilon ^2+(4n-6)V_0^2 
\end{equation}
where $\sigma _\epsilon ^2$ is the variance of single-particle spectrum, and 
$V_0^2$ is the variance of non-diagonal matrix elements of the residual
interaction. Note that in the case of $n<<m$ for low-lying states the
variance of the occupied orbital energies $\sigma _k^2$ is small and the
variance of empty orbital energies is $\sigma _p^2\sim \sigma _\epsilon ^2$.

Thus, the width $\Gamma (E)$ is given by the following expression, 
\begin{equation}
\label{GamSum}\Gamma (E)=2\pi \,\left[ (n-1)V^2\rho _f^{(1)}(E)+V^2\rho
_f^{(2)}(E)\right] 
\end{equation}
The factor $n-1$ appears since for single-particle transitions the summation
in $H_{kp}=\sum_\nu V_{\alpha \nu \rightarrow \gamma \nu }$ is performed
over occupied orbitals. When the ratio $K_2/((n-1)K_1)=(m-n-1)/4$ is larger
than $1$, the two-particle transitions dominate and one can neglect the
differences in $\overline{\omega }$ and $\sigma $ for two-particle and
one-particle transitions. In this case the spreading width is described by
the simple Gaussian form, 
\begin{equation}
\label{GammaG}\Gamma _k(E)\simeq 2\pi (\Delta E)_k^2\frac 1{\sigma _k\sqrt{%
2\pi }}exp\left\{ -\frac{(\tilde E-E_k-\overline{\omega _k})^2}{2\sigma _k^2}%
\right\} 
\end{equation}
where $\tilde E=E-\delta $. Here $(\Delta E)_k^2$ is the variance defined by
Eq.(\ref{H2}) and $\omega _k$ and $\sigma _k$ are close to that for the
two-particle transitions. The maximum of $\rho _f\,(E)$ and $\Gamma (E)$ is
shifted by the value $\left| \overline{\omega _k}\right| $ towards the
center of the spectrum, compared to the maximum of the Breit-Wigner
function. This leads to some distortion of the strength function Eq.(\ref
{FfBW}) and the shape of the eigenstates, which is especially large at the
bottom of the spectrum.

\begin{figure}[htb]
\vspace{-0.5cm}
\begin{center}
\hspace{0.5cm}
\epsfig{file=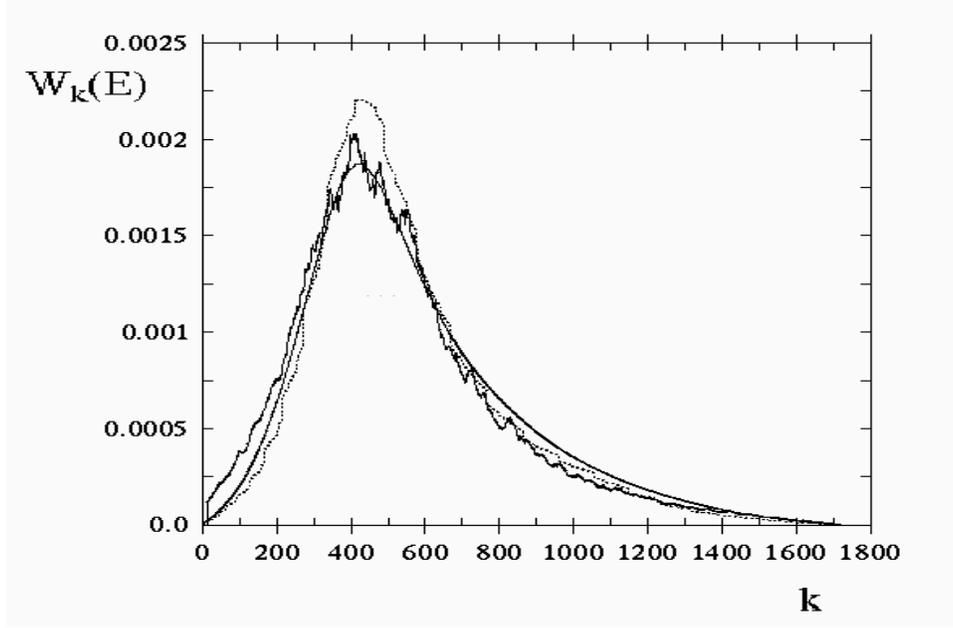,width=5.0in,height=3.3in}
\vspace{0.5cm}
\caption{
Shape of the LDOS (\ref{FfBW}) in the basis representation. Broken
line is the result of numerical diagonalization of the TBRI-matrix. To
reduce fluctuations, the average over $50$ different matrices $H_{ik}$ and
over a number of nearby components has been made. The computation has been
done for $n=6$ particles, $m=13$ orbitals, therefore, the total size of the
matrix is $N=1716.$ The interaction strength is $V_0^2\approx 0.1$ and $%
d_0=1.0$ . Dashed and smooth full curves obtained by computation of Eq.(\ref
{FfBW}) with $\Gamma _k(E)$ given by Eqs.(\ref{Gammaf},\ref{deltaf}) and by
Eq.(\ref{GammaG}) correspondingly.
}
\end{center}
\end{figure}

\begin{figure}[htb]
\vspace{-2.0cm}
\begin{center}
\hspace{-1.5cm}
\epsfig{file=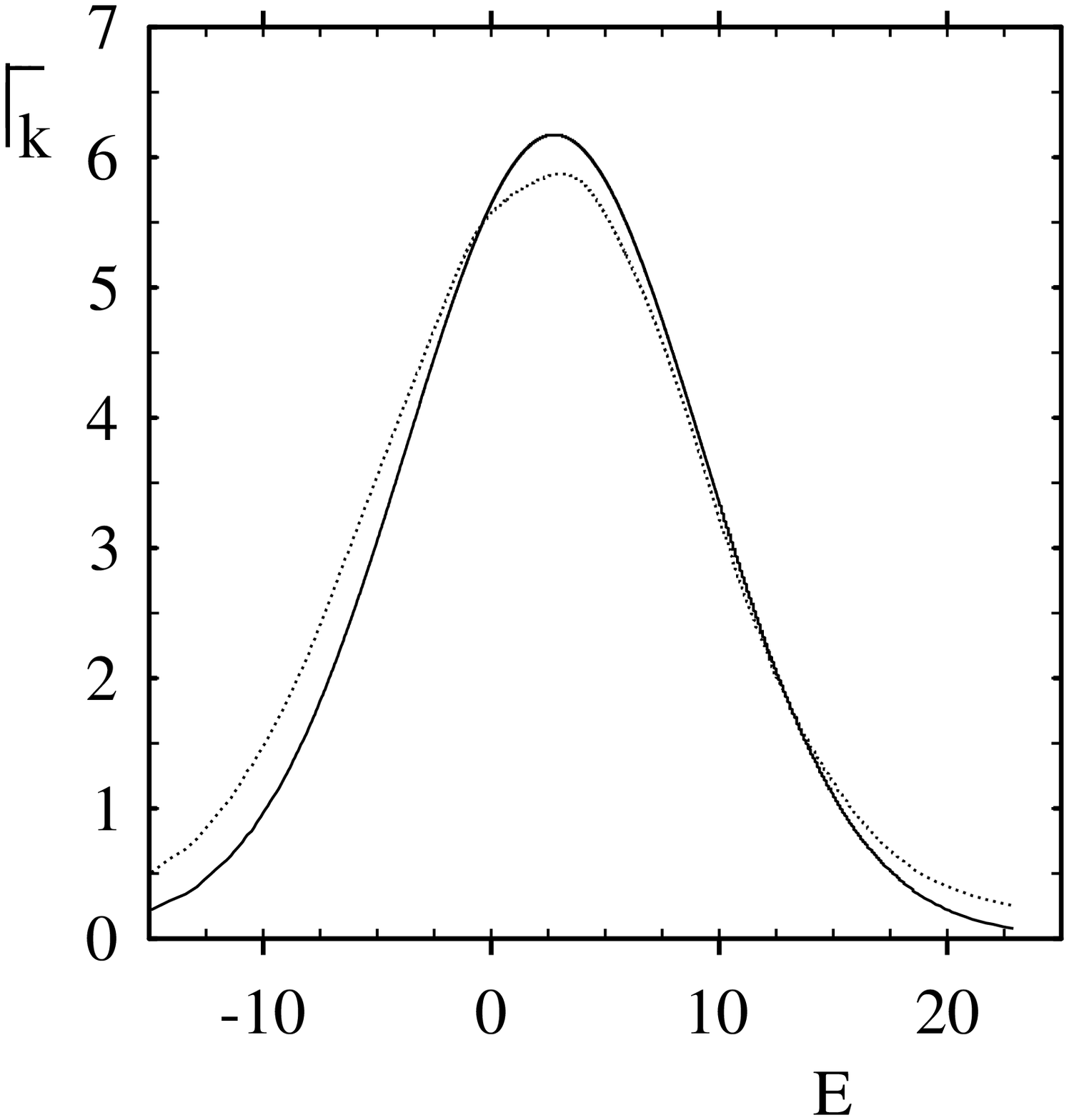,width=5.0in,height=3.3in}
\vspace{-0.5cm}
\caption{
Dependence $\Gamma _k(E)$ is shown for the parameters of Fig.10. Full
curve is the expression (\ref{GammaG}), the dashed curve is the computation
from Eqs.(\ref{Gammaf}, \ref{deltaf}).
}
\end{center}
\end{figure}

The above consideration shows that if the interaction is small, $\Gamma \ll
\sigma _k$ , the strength function has the Breit-Wigner shape with a broad
gaussian envelope described by the dependence $\Gamma _k(E)$ in the
numerator of Eq. (\ref{FfBW}). In fact, such a dependence results in the
correct (finite) variance of the strength function. When the interaction $V$
increases one needs to take into account one more contribution in Eq.(\ref
{Gammaf}) (it was neglected in Eq.(\ref{GammaH})). It increases the width of 
$W_p(E)$ and leads to the estimate $\sigma _k^2\simeq \sigma _2^2+\overline{%
\Gamma _p^2}$. With further increase of interaction, where the shape of $%
W_p(E)$ is close to the Gaussian, one gets $\sigma _k^2\simeq \sigma
_2^2+(\Delta E)_k^2$.

Direct numerical study of the model (\ref{H}) with $n=6$ Fermi-particles and 
$m=13$ orbitals \cite{FI99} confirmed that the above analytical expressions
give quite good description of the shape of the strength function $W_k(E)$
as well as the energy dependence $\Gamma _k(E)\,$ , see Fig.10 and Fig.11.
The size of the Hamiltonian matrix is $N=C_m^n=1716$ and the unperturbed
state $i_0=440\,$was taken.

For the case of quite strong interaction, when $\Gamma \sim \sigma $, the
gaussian variation of $\Gamma (E)$ in the numerator of Eq.(\ref{FfBW})
becomes as important as the variation of the Breit-Wigner energy denominator 
$(E-E_k)^2+(\Gamma /2)^2$. In this case the transition from the Breit-Wigner
type to Gaussian shape of the LDOS (strength function) takes place. However,
still one can use Eqs.(\ref{Gammaf}, \ref{deltaf}) and (\ref{FfBW}) in order
to calculate numerically $\Gamma (E),P_k(E)$ and $F(E,E_k)$, taking $\Gamma $
from Eq.(\ref{GammaG}) with $\sigma _k^2\simeq \sigma _2^2+(\Delta E)_k^2$
as the zero approximation in the right-hand side of Eqs.(\ref{Gammaf}, \ref
{deltaf}).

\subsection{Non-statistical properties of the TBRI-model}

In previous Sections we have considered statistical properties of the
TBRI-model based on chaotic structure of the eigenstates and LDOS. However,
one should be very careful with this approach since due to underlying
correlations in matrix elements of $H$ (see Section 2.1.3),
the approach is not always valid, even if all two-body
random matrix elements are completely random and independent variables. This
fact is due to a two-body nature of interaction and should be taken into
account in some cases. Below we show an example when statistical description
is incorrect (see details in \cite{FGI96}).

Let us consider a single-particle operator 
\begin{equation}
\label{M}\hat M=\sum_{\alpha ,\beta }a_\alpha ^{\dagger }\,a_\beta
\,M_{\alpha \beta }\,=\sum_{\alpha \,,\beta }\rho _{\alpha \beta
}\,\,M_{\alpha \beta } 
\end{equation}
where $a_\alpha ^{\dagger }$ and $a_\beta $ are the creation and
annihilation operators and we have introduced the density matrix operator $%
\rho _{\alpha \beta }=a_\alpha ^{\dagger }a_\beta $ which transfers a
particle from the orbital $\beta $ to the orbital $\alpha $. The matrix
element of $\hat M$ between compound states can be expressed through the
projection of the density matrix into the basis states, 
\begin{equation}
\label{n1n2}\left\langle n_1\right| \hat M\left| n_2\right\rangle
=\sum_{\alpha \beta }M_{\alpha \beta }\left\langle n_1\right| \rho _{\alpha
\beta }\left| n_2\right\rangle =\sum_{\alpha \beta }M_{\alpha \beta }\,\rho
_{\alpha \beta }^{(n_1,n_2)} 
\end{equation}
where 
\begin{equation}
\label{rhon1n2}\rho _{\alpha \beta }^{(n_1,n_2)}\equiv 
\sum_{ij}C_i^{(n_1)}\left\langle i\right| \rho _{\alpha \beta
}\left| j\right\rangle C_j^{(n_2)} 
\end{equation}
is determined by the exact eigenstates only. In what follows, we are
interested in statistical properties of this many-body operator $\hat \rho
(\alpha ,\beta )\,$ for the fixed orbitals $\alpha $ and $\beta $ which, on
the other hand, determines statistical properties of the single-particle
operator $\hat M$ , see Eq.(\ref{M}).

One can see that this operator has zero mean, 
\begin{equation}
\label{meanrho}\overline{\rho (\alpha ,\beta )}=\overline{\left\langle
n_1\right| \rho _{\alpha \beta }\left| n_2\right\rangle }=0 
\end{equation}
if compound eigenstates are truly random.

In general case, the variance of $\hat \rho (\alpha ,\beta )$ which is of
our main interest, has the form,

$$
\overline{\rho ^2(\alpha ,\beta )}=\overline{\left\langle n_1\right| \rho
_{\alpha \beta }\left| n_2\right\rangle \left\langle n_2\right| \rho _{\beta
\alpha }\left| n_1\right\rangle }\,= 
$$
\begin{equation}
\label{M2}\overline{\sum\limits_{i,j,%
\,k,l}C_i^{(n_1)}C_j^{(n_1)}C_k^{(n_2)}C_l^{(n_2)}\left\langle i\right| \rho
_{\alpha \beta }\left| k\right\rangle \left\langle l\right| \rho _{\beta
\alpha }\left| j\right\rangle }=S_d^{(n_{1,}n_2)}+S_c^{(n_{1,}n_2)} 
\end{equation}
Here we separated the diagonal,
\begin{equation}
\label{Sd}S_d^{(n_1n_2)}=\overline{\sum\limits_{ik}\left| C_i^{(n_1)}\right|
^2\left| C_k^{(n_2)}\right| ^2\left| \left\langle i\right| \rho _{\alpha
\beta }\left| k\right\rangle \right| ^2}~, 
\end{equation}
and non-diagonal,

\begin{equation}
\label{Sc}S_c^{(n_{1,}n_2)}=\overline{\sum\limits_{i\neq j,\,k\neq
l}C_i^{(n_1)}C_j^{(n_1)}C_k^{(n_2)}C_l^{(n_2)}\left\langle i\right| \rho
_{\alpha \beta }\left| k\right\rangle \left\langle l\right| \rho _{\beta
\alpha }\left| j\right\rangle }~. 
\end{equation}
contributions to the sum (\ref{M2}) and assumed that eigenstates are real
vectors (note that our matrix $H_{ij}$ is symmetric). Typical shape of the
density matrix is shown in Fig.12. This shape can be compared with the
statistical approach developed in Ref.\cite{DFW77,FKPT88} for a very large
interaction, in the case when the role of the unperturbed part $H_0$ is
neglected (therefore, the influence of the leading diagonal $H_{jj}$ is
small).

\begin{figure}[htb]
\vspace{-0.8cm}
\begin{center}
\hspace{0.4cm}
\epsfig{file=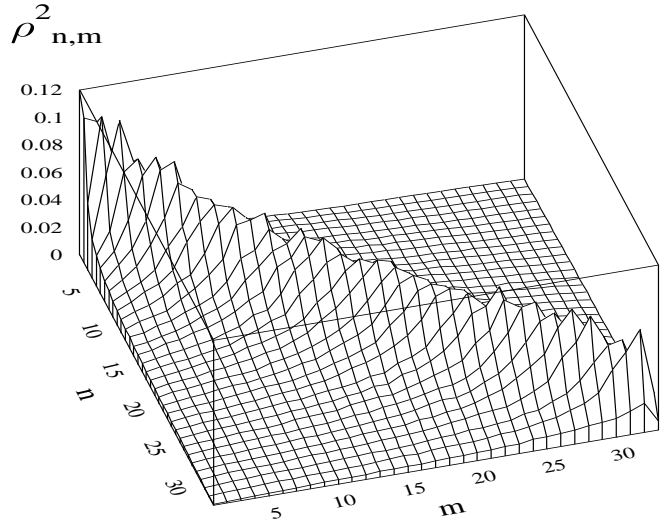,width=5.0in,height=3.3in}
\vspace{-0.5cm}
\caption{
Form of the density matrix (\ref{rhon1n2}) for the TBRI-model with
the parameters of Fig.1-5. The quantity $\rho _{n,m}=\left( \overline{\left(
\rho _{\alpha \beta }^{\left( n_1n_2\right) }\right) ^2}\right) ^{1/2}$ is
computed with the average inside the blocks of size $10\times 10$ (in the
same way as in Fig.2).
}
\end{center}
\end{figure}

As we have discussed in previous Sections, for sufficiently large
interaction compound eigenstates $\left| n\right\rangle \,$of the TBRI-model
may be considered as pseudo-random functions due to a very large number of
components $C_i^{(n)}$ . Therefore, it is natural to expect that the
non-diagonal part $S_c^{(n_{1,}n_2)}\,$is zero and the variance is
essentially determined by the diagonal term $S_d^{(n_1,n_2)}\,$ which can be
described statistically (for this statistical approach see Refs.\cite
{F94,FGGK94,FV93}). This assumption has been used in the previous
calculations of matrix elements between compound states in \cite
{F94,FGGK94,FV93,SF82}. However, recently it was shown \cite{FGI96} that in
many-body system these two terms are of the same order, $S_c\sim S_d$, even
for completely random two-body interaction. In order to show this very
unexpected effect, in Ref.\cite{FGI96} direct computations of the terms $S_c$
and $S_d$ have been performed for the TBRI-model, see Figs.13-14. One can
see that the data reveal a systematic difference between the diagonal
approximation and exact expression (\ref{M2}). In particular, Fig.14 shows
that non-diagonal term $S_c$ is of the same order as $S_d$ which clearly
indicates the presence of correlations.

Below, we show how these correlations emerge in the non-diagonal term $S_c$
(for more details see Ref.\cite{FGI96}). First, note that for a given $i$
the sum over $k$ in Eq. (\ref{Sd}) for $S_d$ contains only one term, for
which $|k\rangle =a_\beta ^{\dagger }a_\alpha |i\rangle \equiv |i^{\prime
}\rangle $, determined by transferring one particle from the orbital $\alpha 
$ to the orbital $\beta $ in the state $|i\rangle $ (hereafter we use the
notation $i^{\prime }$ to mark such states). Accordingly, the index $i$ runs
over those states in which $\alpha $ is occupied and $\beta $ is vacant. For
such $i$ and $i^{\prime }$ the matrix element $\left\langle i\right| \rho
_{\alpha \beta }\left| i^{\prime }\right\rangle \,=1$, otherwise, it is
zero. Therefore, in fact, the sum in (\ref{Sd}) is a single sum, with a
number of items less than $N$, 
\begin{equation}
\label{SSd}S_d^{(n_1n_2)}=\overline{{\sum_i}^{\prime }\left|
C_i^{(n_1)}\right| ^2\left| C_{i^{\prime }}^{(n_2)}\right| ^2} 
\end{equation}
where the sum ${\sum_i}^{\prime }$ runs over the specified $i$. Analogously,
Eq. (\ref{Sc}) can be written as the double sum over $i$ and $j$ specified
as above, 
\begin{equation}
\label{SSc}S_c^{(n_1n_2)}=\overline{{\sum\limits_{i\neq j}}^{\prime \prime
}C_i^{(n_1)}C_j^{(n_1)}C_{i^{\prime }}^{(n_2)}C_{j^{\prime }}^{(n_2)}}, 
\end{equation}
where $j^{\prime }$ is a function of $j$, $|j^{\prime }\rangle =a_\beta
^{\dagger }a_\alpha |j\rangle $. Note that the energies of the basis states
and their primed partners are connected as $E_{i^{\prime }}-E_i=\epsilon
_\beta -\epsilon _\alpha =E_{j^{\prime }}-E_j$.

\begin{figure}[htb]
\vspace{-0.8cm}
\begin{center}
\hspace{-1.0cm}
\epsfig{file=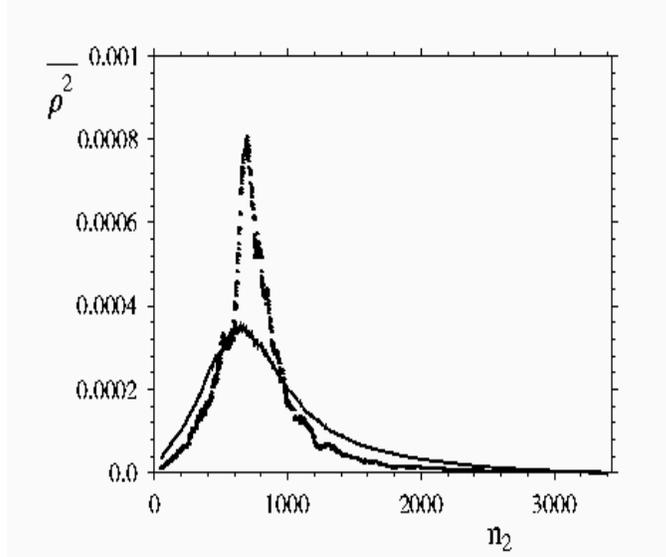,width=3.0in,height=3.5in,angle=-90}
\vspace{-0.0cm}
\caption{
Mean square matrix element (\ref{M2}) calculated in the TBRI-model for $%
n=7$ particles and $m=14$ orbitals, $\alpha =7$, $\beta =8$, as a function
of the eigenstate $n_2$ for $n_1=575$ (total size of the matrix is $N=3432)$%
. Dots correspond to the sum $S_d+S_c$ while the solid line represents the
diagonal contribution $S_d$ only [see (\ref{Sd})].
}
\end{center}
\end{figure}

\begin{figure}[htb]
\vspace{-1.2cm}
\begin{center}
\hspace{-1.0cm}
\epsfig{file=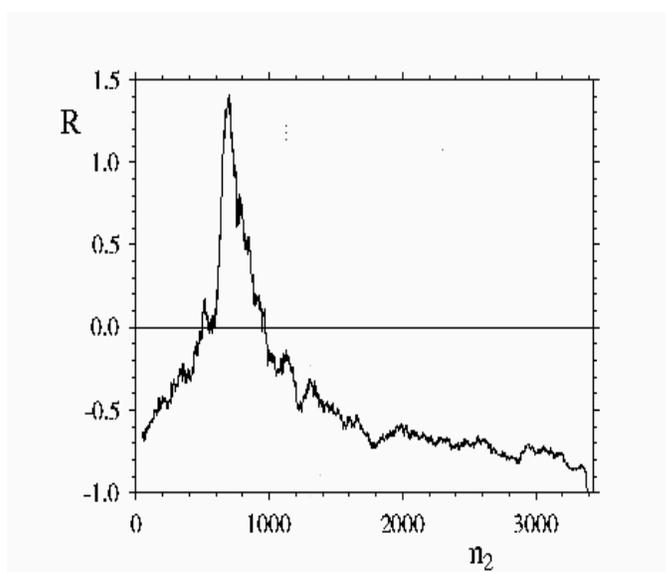,width=3.0in,height=3.5in,angle=-90}
\vspace{-0.0cm}
\caption{
Ratio $R=S_c/S_d$ of the correlation contribution to the diagonal
contribution for the same parameters as in Fig.13.
}
\end{center}
\end{figure}

One can expect that maximal values of the sum (\ref{SSd}) and (\ref{SSc})
correspond to the terms for which $C-$components are principal components of
the eigenstates. This means that mean square of the matrix element $
\overline{\left| \left\langle n_1\right| \rho _{\alpha \beta }\left|
n_2\right\rangle \right| ^2}$ is maximal when the operator $\rho _{\alpha
\beta }$ couples the principal components of the state $\left|
n_1\right\rangle $ with those of $\left| n_2\right\rangle $, i.e. for $%
E^{(n_1)}-E^{(n_2)}\approx \omega _{\alpha \beta }$ $\equiv \epsilon _\alpha
-\epsilon _\beta $. Far from the maximum ($\left| E^{(n_1)}-E^{(n_2)}-\omega
_{\alpha \beta }\right| >\Gamma $) a principal component of one state, say, $%
n_1$, is coupled to a small component $k$ of the other state $n_2$ ($\left|
E_k-E^{(n_2)}\right| >\Gamma )$. The latter case is simpler to consider
analytically, since the admixture of a small component in the eigenstate can
be found via the standard perturbation theory. This approach reveals the
origin of the correlations in the sum $S_c$, Eq.~(\ref{SSc}). For example,
if $C_j^{(n_1)}$ is a small component of the eigenstate $n_1$, then it can
be expressed as a perturbation theory admixture to the principle components.
If the component $C_i^{(n_1)}$ is one of the latter, then there is a term in
the sum (\ref{SSc}), which is proportional to the principal component
squared, $\left| C_i^{(n_1)}\right| ^2$.

Based on this consideration, in Ref.\cite{FGI96} was found that far from the
maximum, $\left| E^{(n_2)}-E^{(n_1)}-\omega _{\beta \alpha }\right| >\Gamma $%
, the non-diagonal terms read 
\begin{equation}
\label{Sc34}S_c^{(n_1n_2)}\approx -\frac 2{\left( E^{(n_2)}-E^{(n_1)}-\omega
_{\beta \alpha }\right) ^2}\widetilde{\sum\limits_{i,j\prime }\prime \prime }%
\left| C_i^{(1)}\right| ^2\left| C_{j^{\prime }}^{(2)}\right| ^2\overline{%
H_{i^{\prime }j^{\prime }}H_{ij}} 
\end{equation}
A similar calculation of the diagonal sum $S_d^{(n_1n_2)}$, Eq. (\ref{Sd}),
yields%
$$
S_d^{(n_1n_2)}\approx \frac 1{\left( E^{(n_2)}-E^{(n_1)}-\omega _{\beta
\alpha }\right) ^2} 
$$
\begin{equation}
\label{Sd34}\times \left\{ \widetilde{\sum\limits_i\prime }\widetilde{%
\sum\limits_{j\prime }}\left| C_i^{(n_1)}\right| ^2\left| C_{j^{\prime
}}^{(n_2)}\right| ^2\overline{H_{i^{\prime }j^{\prime }}^2}+\widetilde{%
\sum\limits_i}\widetilde{\sum\limits_{j\prime }\prime }\left|
C_i^{(n_1)}\right| ^2\left| C_{j^{\prime }}^{(n_2)}\right| ^2\overline{%
H_{ij}^2}\right\} 
\end{equation}

Let us estimate the relative magnitudes of $S_d$ and $S_c$. First, we
consider the case when $|i\rangle $ and $|j\rangle $ differ by two orbitals, 
$|j\rangle =a_{\mu _2}^{\dagger }a_{\mu _1}a_{\nu _2}^{\dagger }a_{\nu
_1}|i\rangle ,$ in this case $H_{ij}=V_{\nu _1\mu _1\nu _2\mu _2}$. Since
the basis states $|i^{\prime }\rangle $ and $|j^{\prime }\rangle $ must
differ by the same two orbitals, we have $H_{i^{\prime }j^{\prime }}=V_{\nu
_1\mu _1\nu _2\mu _2}=H_{ij}$ (note that $\nu _1,\mu _1,\nu _2,\mu _2\neq
\alpha ,\beta $, since both states $|i\rangle $ and $|j\rangle $ contain $%
\alpha $ and do not contain $\beta $, whereas $|i^{\prime }\rangle $ and $%
|j^{\prime }\rangle $ contain $\beta $ and do not contain $\alpha $).
Therefore, the averages over the non-zero matrix elements between such pairs
of states give $\overline{H_{ij}H_{i^{\prime }j^{\prime }}}=\overline{%
H_{ij}^2}=\overline{H_{i^{\prime }j^{\prime }}^2}=V_0^2$.

Now, let us consider the case when $|i\rangle $ and $|j\rangle $ differ by
one orbital $|j\rangle =a_{\nu _2}^{\dagger }a_{\nu _1}|i\rangle $ only. In
this case the Hamiltonian matrix elements are sums of the $n-1$ two-body
matrix elements, see Eqs.(\ref{corr1}) and (\ref{corr1a}). As was shown in
Section 2.1.3, the sums of $n-2$ terms in $H_{ij}$ and $H_{i^{\prime
}j^{\prime }}$ coincide and the difference is due to the one term only
(orbital $\alpha $ is replaced by the orbital $\beta $). Thus, 
\begin{equation}
\label{HHH}\overline{H_{ij}H_{i^{\prime }j^{\prime }}}=(n-2)V^2~,\,\,\,\,\,%
\,\,\,\overline{\left( H_{ij}\right) ^2}=\overline{\left( H_{i^{\prime
}j^{\prime }}\right) ^2}=(n-1)V^2~ 
\end{equation}
where we took into account that $\overline{V_{\kappa \lambda \mu \nu
}V_{\kappa _1\lambda _1\mu _1\nu _1}}=V^2\delta _{\kappa \kappa _1}\delta
_{\lambda \lambda _1}\delta _{\mu \mu _1}\delta _{\nu \nu _1}$.

The contributions of one-particle and two-particle transitions in Eqs. (\ref
{Sc34}) and (\ref{Sd34}) representing $S_c$ and $S_d$ respectively, are
determined by the numbers of such transitions allowed by the corresponding
sums. For the single-prime sums in Eq. (\ref{Sd34}) these numbers are
proportional to $K_1$ and $K_2$, Eq. (\ref{K123}). In the double-prime sum
in Eq. (\ref{Sc34}) these numbers are proportional to $\tilde K_1$ and $%
\tilde K_2$, the numbers of the two-body and one-body transitions $%
i\rightarrow j$, in the situation when one particle and the two orbitals ($%
\alpha $ and $\beta $) do not participate in the transitions. These numbers
can be obtained from Eq. (\ref{K123}) if we replace $n$ by $n-1$, and $m$ by 
$m-2$, so that $\tilde K_1=(n-1)(m-n-1)$, $\tilde
K_2=(n-1)(n-2)(m-n-1)(m-n-2)/4$. Finally one can obtain that for $%
|E^{(n_2)}-E^{(n_1)}-\omega _{\beta \alpha }|>\Gamma $ the contribution of
the correlation term to the variance of the matrix elements of $\rho
_{\alpha \beta }$ can be estimated in the ratio as 
\begin{equation}
\label{final}R\equiv \frac{S_c}{S_d}=-\frac{(n-2)\tilde K_1+\tilde K_2}{%
(n-1)K_1+K_2}=-\frac{(n-2)(m-n-1)(m-n+2)}{n(m-n)(m-n+3)}. 
\end{equation}
This equation shows that for $n=2$ we have $S_c=0$, which is easy to check
directly, since in this case $\overline{H_{i^{\prime }j^{\prime }}H_{ij}}=0$
. For $n>2$ the correlation contribution $S_c$ is negative in the tails of
the LDOS. This means that it indeed suppresses the transition amplitudes
off-resonance (see Figs.13-14). For $n,m-n\gg 1$ the ratio $R$ approaches
its limit value $-1$. It is easy to obtain from Eq. (\ref{final}) that for $%
m-n\gg 1$ 
\begin{equation}
\label{final1}\frac{S_d+S_c}{S_d}=1+R\simeq \frac{2m}{n(m-n)}~. 
\end{equation}

Thus, surprisingly, the role of the correlation contribution increases with
the number of particles. This result is supported by numerical data reported
in Ref.\cite{FGI96}. In Fig.14 one can see that the suppression of the
matrix elements $\overline{\rho ^2(\alpha ,\beta )}$ due to the correlation
term at the tails, is quite strong, numerical ratio is $R\approx -0.7$ vs. $%
R=-0.55$ obtained from Eq.(\ref{final}); this should be compared with the
case $n=4,m=11,N=330$ for which numerical value is $R\approx -0.45,$ see
details in \cite{FGI96}. The correlation contribution should be even more
important in compound nuclei, where $N\sim 10^5$ . This case can be modeled
by the parameters $n=10$, $m=20$; then we have $R=-0.66$, or, equivalently, $%
(S_d+S_c)/S_d=0.34$, which means that the correlations suppress the squared
element $\rho^2$ between compound states by a factor of $3$ (far from its
maximum).

It is worth emphasizing that the existence of correlations due to the
perturbation theory admixtures of small components to the chaotic
eigenstates, which leads to a non-zero value of $S_c$ (\ref{SSc}), is indeed
non-trivial. For example, if one examines the sum of Eq. (\ref{SSc}) as a
function of $i$ and $j$, it would be hard to guess that the sum itself is
essentially non-zero, since positive and negative values of $\xi _{ij}\equiv
C_i^{(n_1)}C_j^{(n_1)}C_{i^{\prime }}^{(n_2)}C_{j^{\prime }}^{(n_2)}$ seem
to be equally frequent, and have roughly the same magnitude, see Fig.15.
However, in spite of apparent random character of the terms $\xi _{ij}$ ,
its mean value turns out to be non-zero and is of the same order as the
diagonal term $S_d$ . Since $\sum_{n_1}S_c^{(n_1n_2)}=%
\sum_{n_2}S_c^{(n_1n_2)}=0$ (see below), the suppression of $\overline{\rho
^2(\alpha ,\beta )}$ at the tails should be accompanied by correlational
enhancement of the matrix elements near the maximum (for $\left|
E^{(2)}-E^{(1)}-\omega _{\beta \alpha }\right| <\Gamma $).

Thus, we come to the important conclusion: even for a random two-body
interaction, the correlations produce some sort of a ``{\it correlation
resonance}'' in the distribution of the squared matrix elements $\overline{%
\rho ^2(\alpha ,\beta )}$. One should note that this increase of the
correlation effects in the matrix elements of a perturbation can be
explained by the increased correlations between the Hamiltonian matrix
elements when the number of particles and orbitals increases ($N/n\propto
e^n)$.

In a similar way one can estimate the size of the correlation contribution $%
S_c$ near the maximum of the $M^2$ distribution (at $\left|
E^{(n_2)}-E^{(n_1)}-\omega _{\beta \alpha }\right| <\Gamma )$ \cite{FGI96}. 
\begin{equation}
\label{estimate}\frac{S_d+S_c}{S_d}=1+R_m=2-(1+R_t)\simeq 2\left[ 1-\frac
m{n(m-n)}\right] ~. 
\end{equation}
Comparing the values of the ratio $S_c/S_d$ at the maximum and at the tail
in Fig.13 ($n=7,~m=14$) , one can see that indeed, $R_m\approx -R_t$. For
larger $n$ and $m$ the correlation enhancement factor asymptotically reaches
its maximal value of $2$. This numerical example shows the enhancement of $
\overline{\rho ^2(\alpha ,\beta )}$ with respect to $S_d$ at the maximum
even greater in size than that predicted by Eq. (\ref{estimate}). This is
not too surprising since in Eqs. (\ref{BW})--(\ref{estimate}) we estimated
the average value of $R_m$ over an interval $\Delta E\simeq \Gamma $ around
the maximum rather than the peak value at the maximum.

\begin{figure}[htb]
\vspace{-0.5cm}
\begin{center}
\hspace{0.4cm}
\epsfig{file=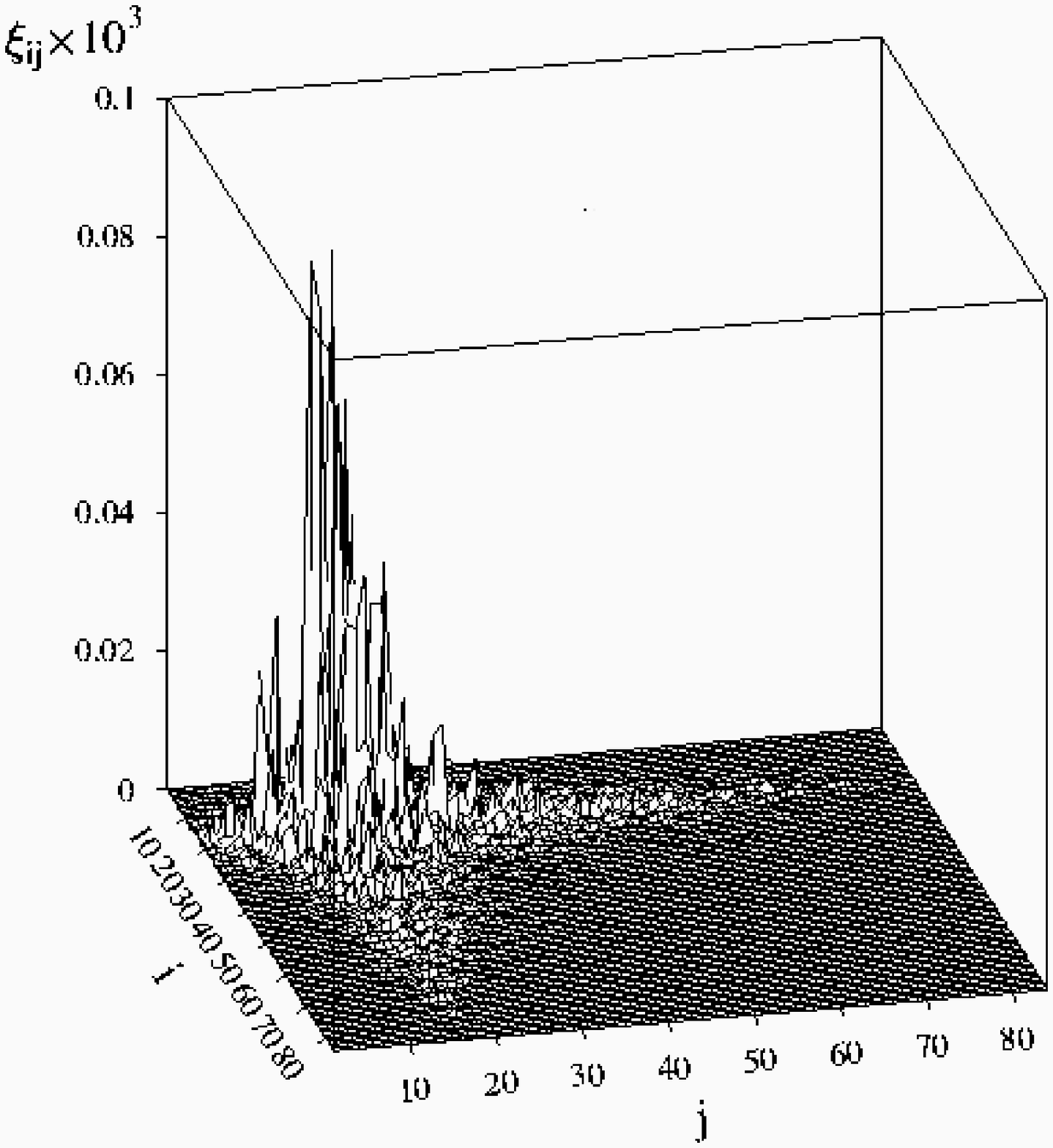,width=4.5in,height=3.7in}
\vspace{-0.5cm}
\end{center}
\end{figure}

\begin{figure}[htb]
\vspace{-2.8cm}
\begin{center}
\hspace{0.4cm}
\epsfig{file=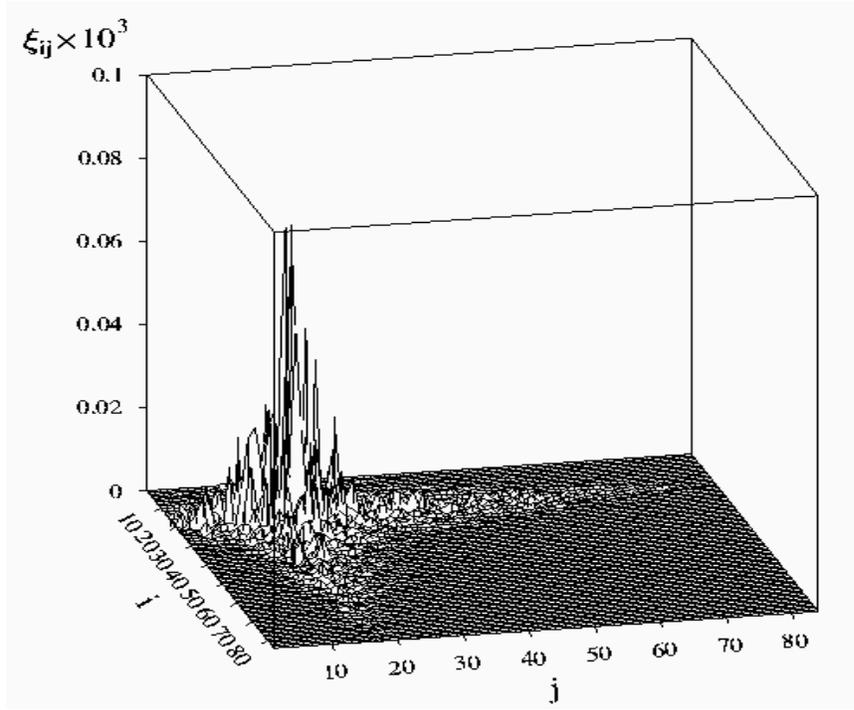,width=4.5in,height=3.7in}
\vspace{0.5cm}
\caption{
The distribution of the terms $\xi
_{ij}=C_i^{(n_1)}C_j^{(n_1)}C_{i^{\prime }}^{(n_2)}C_{j^{\prime }}^{(n_2)}$
of the sum (\ref{SSc}) for $n_1=55$, $n_2=66$, obtained in the TBRI-model
for the same set of parameters as in Figs.1-5, averaged over $N_g=100$
realizations of the interaction. Indices $i$ and $j$ in the figure run over
those $84$ components in which $\alpha $ is occupied and $\beta $ is vacant.
(a) on the top: positive values, (b) on the bottom:
negative values (absolute values).
}
\end{center}
\end{figure}

\section{Thermalization and onset of chaos}

\subsection{Distribution of occupation numbers}

Let us now come back to the distribution of occupation numbers defined by
Eq.(\ref{ns}). It gives the probability that one of $n$ particles occupies
an orbital $s$ specified by the one-particle state $\left| s\right\rangle $
, for the fixed exact (compound) state $\left| i\right\rangle \,$ .
According to this expression, this probability can be found by projecting
the state $\left| i\right\rangle \,$ onto the basis of unperturbed states,
for which the relation between the positions of all particles in the
single-particle basis and specific many-particle basis state is known by the
construction of the latter. One can see that the probability $%
n_s=n_s(E^{(i)})$ is the sum of probabilities over number of basis states
which construct the exact state. For Fermi-particles, only one particle can
occupy an orbital, this is why the occupation number $n_s^{(k)}=0$ or $1$ .

It is clear that for chaotic eigenstates the $n_s-$distribution is a
fluctuating function of the total energy $E=E^{(i)}$ of a system, due to
fluctuations of the components $C_k^{(i)}$ of eigenstates.
In order to obtain a smooth dependence, one should make an average over a
small energy window centered at $E^{(i)}$ , which is in the spirit of
conventional statistical mechanics for systems in the contact with the
thermostat. In fact, such an average is a kind of microcanonical averaging
since it is done for the fixed total energy $E$ of a system. Therefore, in
what follows, by the $n_s-${\it distribution} we assume the averaged
distribution, 
\begin{equation}
\label{nsAv}n_s(E)=\sum\limits_k\overline{\left| C_k^{(i)}\right| ^2\,}%
\left\langle k\right| \hat n_s\left| k\right\rangle
=\sum\limits_kF(E_k,E^{(i)})\overline{\,}\left\langle k\right| \hat
n_s\left| k\right\rangle 
\end{equation}
where the $F-$function discussed in previous Section is used.

We are going to show that the $n_s-$ distribution plays essential role in
the statistical approach to finite systems of interaction particles. Our
interest to this quantity is of two-fold. First, the knowledge of the
distribution of occupation numbers gives the possibility to calculate mean
value of any single particle operator $\left\langle M\right\rangle
=\sum_sn_sM_{ss}$ . Here $M_{ss}$ are diagonal matrix elements of a
single-particle operator $\hat M$ , which very often can be found easily
since they refer single-particle physics. For quantum systems with complex
behavior, the non-trivial part is the $F-$function which absorbs the result
of the two-body interaction between many basis states. The important point
is that in order to find the distribution of occupation numbers, there is no
need to know exactly eigenstates of the system. Instead, the
shape of eigenstates in the energy representation is needed, which is
defined by the $F-$function. Thus, if we know this function and properties
of the unperturbed system, one can relate statistical properties of chaotic
systems to single-particle quantities.

Moreover, the variance of the distribution of non-diagonal elements of $M$ ,
describing transition amplitudes between ``chaotic'' compound states due to
a weak external perturbation, can be also expressed through to the
occupation numbers $n_s$ . This variance is important from experimental
point of view, for example, for the estimate of an enhancement of a weak
interaction (which refers,
 for example, the parity violation in atoms and nuclei) , see
details in \cite{SF82,FGGK94,FGI96,FGGP99}.

Second, the form of the distribution of occupation numbers is interesting
itself. One of important questions which arises in this respect, is whether
the standard Fermi-Dirac and Bose-Einstein distributions occur for isolated
systems of finite number of particle. If they occur, then, under what
conditions and how they can be described? One can suggest that the role of
the interaction between particles is crucial since this is the only reason
to result in a complex behavior (chaos) of a system, and the latter is the
mechanism for the statistical equilibrium. Another non-trivial question
relates to the meaning of temperature for isolated systems. The study of the 
$n_s-$distribution gives a new insight on these and other problems.

For the analytical treatment it is convenient to represent the $n_s-$%
distribution in the following form, 
\begin{equation}
\label{nalpha}n_s(E)=\frac{\sum\limits_kn_s^{(k)}\,\tilde F(E_k-E\,)}{%
\sum\limits_k\,\tilde F(E_k-E)} 
\end{equation}
Here the function $\tilde F(E_k-E)$ is the part of the $F-$function, which
non-trivially depends on the difference $E_k-E\,$ between the unperturbed
energy $E_k$ only. Namely, we omitted the normalization term $f\,(E)$ since
the summation in Eq.(\ref{nalpha}) runs over the unperturbed energy, $%
F(E_k,E)=f\,(E)\,\tilde F(E_k-E\,)$ . Thus, the denominator appears due to
the normalization. This form (\ref{nalpha}) allows one to introduce a kind
of the {\it partition function}, 
\begin{equation}
\label{ZZ}Z(E)=\sum_k\tilde F(E_k-E) 
\end{equation}
which is entirely determined by the shape of chaotic eigenfunctions.

The above expression (\ref{nalpha}) gives a possibility of the statistical
description of complex systems. Indeed, as was mentioned above, the shape of
the $F-$ function has universal features and can be often described
analytically. Therefore, in practice there is no need to diagonalize huge
Hamiltonian matrix of a many-body system in order to find statistical
averages. We would like again to stress that the summation in (\ref{nalpha})
is carried out over unperturbed energies $E_k$ defined by the mean field,
rather than over the energies of exact eigenstates in the standard canonical
distribution. As a result, the distribution of occupation numbers can be
derived analytically (see below) even for few interacting particles, in a
situation when the standard Fermi-Dirac distribution does not occur.

\subsection{Microcanonical vs. canonical distribution}

Let us now compare the $n_s-$distribution (\ref{nalpha}) with occupation
numbers given by the standard {\it canonical distribution} \cite{FI97}, 
\begin{equation}
\label{gibbs}n_s(T)=\frac{\sum\limits_in_s^{(i)}\,\exp (-E^{(i)}/T)}{%
\sum\limits_i\exp (-E^{(i)}/T)} 
\end{equation}
Here $T$ is the temperature of a heat bath and the index $i$ stands for
exact eigenstates. The important difference between the $n_s-$ distribution (%
\ref{nalpha}) and the canonical distribution (\ref{gibbs}) is that in Eq. (%
\ref{nalpha}) the occupation numbers are calculated for the fixed total
energy $E$ of a system unlike the fixed temperature $T$ in Eq.(\ref{gibbs}).
However, both expressions can be compared with each other using the relation
between the energy $E$ and the temperature $T$ , 
\begin{equation}
\label{energy}E=\left\langle E\right\rangle _T=\frac{\sum\limits_iE^{(i)}\,%
\exp (-E^{(i)}/T)}{\sum\limits_i\exp (-E^{(i)}/T)} 
\end{equation}
One can show that the canonical distribution corresponds to the average of
the ``{\it microcanonical}'' $n_s-$ distribution over some energy interval $%
\Delta _T$ . To demonstrate this, let us substitute $n_s^{(i)}$ and $
\overline{\left| C_k^{(i)}\right| ^2}$ from Eqs.(\ref{ns},\ref{Fdef}) into
Eq.(\ref{gibbs}) and replace the summation over $i$ by the integration over $%
\rho (E^{(i)})\,dE^{(i)}$ where $\rho (E^{(i)})$ is the density of exact
energy levels, 
\begin{equation}
\label{Ffi}\sum_in_s^{(i)}\exp \left( -E^{(i)}/T\right) \approx \int
n_s^{(i)}\rho (E^{(i)})\exp \left( -E^{(i)}/T\right) dE^{(i)}\approx 
\end{equation}
$$
\sum_kn_s^{(k)}\int F_k^{(i)}(E_k,E^{(i)})\rho (E^{(i)})\exp \left(
-E^{(i)}/T\right) dE^{(i)}=\sum\limits_kn_s^{(k)}\,F(T,E_k) 
$$
Here the function $F(T,E_k)$ is the canonical average of $F_k^{(i)}$ , 
\begin{equation}
\label{FT}F(T,E_k)=\int F_k^{(i)}\Phi _T(E^{(i)})\,dE^{(i)} 
\end{equation}
with another ``canonical (thermal) averaging'' function, 
\begin{equation}
\label{fi}\Phi _T(E)=\rho (E)\exp \left( -E/T\right) 
\end{equation}
As a result, one can transform the canonical distribution (\ref{gibbs}) into
the form similar to the $n_s-$ distribution (\ref{nalpha}), 
\begin{equation}
\label{nsT}n_s(T)=\frac{\sum\limits_kn_s^{(k)}\,F(T,E_k)}{%
\sum\limits_k\,F\,(T,E_k)} 
\end{equation}

This distribution can be used for the calculation of the occupation numbers
and other mean values in a quantum dot which is in thermal equilibrium with
an environment (with no particle exchange).

In many-body systems with large number of particles the function $\Phi _T(E)$
has a very narrow maximum since the density of states $\rho (E^{(i)})\,$
typically grows very fast. The position $E_m$ of its maximum is defined by
the expression 
\begin{equation}
\label{Tt}\frac{d\ln \rho (E)}{dE}=\frac 1T 
\end{equation}
and the width is given by 
\begin{equation}
\label{deltaT}\Delta _T=\left| \frac{d^2\ln \,\rho (E)}{dE^2}\right| ^{-1/2} 
\end{equation}
Let us consider the TBRI-model (\ref{H}) for which the density of states is
known to be described by the Gaussian, 
\begin{equation}
\label{rho}\rho (E)=\frac 1{\sigma \sqrt{2\pi }}\exp \left( -\frac{\left(
E-E_c\right) ^2}{2\sigma ^2}\right) 
\end{equation}
with $E_c$ and $\sigma ^2$ as the center and the variance of the spectrum.
This allows easily to find the form of $\Phi _T(E)$ which also has the
gaussian form, 
\begin{equation}
\label{Fform}\Phi _T(E)=\frac 1{\sigma \sqrt{2\pi }}\exp \left( -\frac{%
\left( E-E_m\right) ^2}{2\sigma ^2}\right) 
\end{equation}
with 
\begin{equation}
\label{Em}E_m=E_c-\frac{\sigma ^2}T 
\end{equation}
One can see that the width $\Delta _T$ of the thermal averaging function is
equal to the gaussian width of the spectrum, $\Delta _T=\sigma $ . In Ref.%
\cite{FI97} it was argued that for large number of particles both widths $%
\Delta _T$ and $\Delta E$ are much smaller than the typical energy interval, 
$\sigma /\left| E-E_c\right| \sim 1/\sqrt{n}$ . Therefore, for large number
of particles the function $\Phi _T$ can be regarded as the delta-function at 
$E=E_m$ and the $n_s-$ distribution is close to the canonical one, see Eq.(%
\ref{FT}).

However, the canonical distribution (\ref{gibbs}) is not correct when
describing isolated systems with small number of particles, instead, one
should use the $n_s-$ distribution (\ref{nalpha}), see details below and 
in \cite{FIC96,FI97}.

\subsection{Transition to the Fermi-Dirac distribution}

It is naturally to expect that for a very large number of particles the
standard Fermi-Dirac distribution arises from the $n_s-$ distribution (\ref
{nalpha}). Below we reproduce the derivation given in Ref.\cite{FI97}. By
splitting the sum in two parts, which corresponds to the separate summation
over $n_s=0$ and $n_s=1$ , one can represent the expression (\ref{nalpha})
in the form 
\begin{equation}
\label{nsZ}n_s(E)=\frac{0+Z_s(n-1,E-\tilde \epsilon _s)}{Z_s(n-1,E-\tilde
\epsilon _s)+Z_s(n,E)}=\frac 1{1+\frac{Z_s(n,E)}{Z_s(n-1,E-\tilde \epsilon
_s)}} 
\end{equation}
where two ``partial'' partition functions $Z_s(n,E)$ and $Z_s(n-1,E-\tilde
\epsilon _s)$ are introduced. For the first function the summation is taken
over all single-particle states of $n$ particles with the orbital $s$
excluded, $Z_s(n,E)=\sum\nolimits_k^{\prime }\tilde F(E_k-E)$.
Correspondingly, the sum in $Z_s(n-1,E-\tilde \epsilon _s)$ is taken over
the states of $n-1$ particles with the orbital $s$ excluded. The latter sum
results from the terms for which the orbital $s$ is filled $(n_s=1)$ , thus,
one should add the energy $\tilde \epsilon _s\equiv E_k(n)-E_k(n-1)$ of this
orbital to the energy $E_k(n-1)$ of the basis state $\left| k\right\rangle $
defined by $n-1$ particles. Since the $F-$ function mainly depends on the
difference $E_k+\tilde \epsilon _s-E$ , the adding term $\tilde \epsilon _s$
to $E_k(n-1)$ is the same as its subtraction from the total energy $E$ .
Note, that this term is defined by

\begin{equation}
\label{eps}\tilde \epsilon _s=\epsilon _s+\sum\limits_{p\neq
s}u_{sp}n_p^{(k)} 
\end{equation}
where $\epsilon _s$ is the energy of a single-particle state and $u_{sp}$ is
the diagonal matrix element of the two-body residual interaction. By taking $%
\tilde \epsilon _s$ independent of $k$ we assume that the averaging over the
basis states near the energy $E$ is possible, in fact, this is equivalent to
a local (at a given energy) mean field approximation. One should stress that
this approximation is important when comparing the simple TBRI-model (\ref{H}%
) with realistic systems. For example, for the Ce atom there are orbitals
from different open sub-shells (e.g. $4f$ and $6s$ ) which are quite close
in energies, however, they have very different radius. As a result, the
Coulomb interaction between the corresponding electrons is very different 
\cite{FGGP98}. In this case the interaction terms in Eq.(\ref{eps}) strongly
depend on the occupation numbers of other particles. As a result, the
equilibrium distribution for occupation numbers $n_s$ is very different from
the Fermi-Dirac distribution \cite{FGGP98}. However, the original $n_s-$
distribution (\ref{nalpha}) for occupation numbers is valid and gives
correct result \cite{PFGG97}. In other cases like random two-body interaction
model \cite{FIC96,FGI96,FI97} or nuclear shell model \cite
{HZB95,ZBHF96}, or the atom of gold \cite{FGG99}, such a local mean
field approximation is quite accurate and results in the FD-distribution.

For large number $n\gg 1$ of particles distributed over $m\gg 1$ orbitals,
the dependence of $Z_s\,$ on $n$ and $\tilde \epsilon _s$ is very strong
since the number of terms $N$ in the partition function $Z_s\,$ is
exponentially large, $N=\frac{m!}{(m-n)!n!}$ . Therefore, to make the
dependence on arguments smooth, one should consider $\ln \,Z_s$ instead of $%
Z_s$ . In this case one can obtain 
\begin{equation}
\label{ln}\ln \,Z_s(n-\Delta n,E-\tilde \epsilon _s)\approx \ln
\,Z_s(n,E)-\alpha _s\,\Delta n\,-\beta _s\tilde \epsilon _s 
\end{equation}
$$
\alpha _s=\frac{\partial \ln \,Z_s}{\partial n};\,\,\,\,\,\,\beta _s=\frac{%
\partial \ln \,Z_s}{\partial E};\,\,\,\,\,\,\Delta n=1 
$$
This leads to the distribution of the Fermi-Dirac type, 
\begin{equation}
\label{FDtype}n_s=\frac 1{1+\exp (\alpha _s+\beta _s\tilde \epsilon _s)} 
\end{equation}
If the number of substantially occupied orbitals in the definition of $Z_s\,$
is large, the parameters $\alpha _s$ and $\beta _s$ are not sensitive as to
which particular orbital $s$ is excluded from the sum and one can assume $%
\alpha _s=\alpha \equiv -\mu /T,\,\,\beta _s=\beta \equiv 1/T$ as in the
standard derivation of the Fermi-Dirac distribution for systems in contact
with thermostat. Therefore, the chemical potential $\mu $ and temperature $T$
can be found from the conditions of fixed number of particles and fixed
energy, 
\begin{equation}
\label{eqs}\sum\limits_sn_s=n;\,\,\,\,\,\sum\limits_s\epsilon
_sn_s+\sum\limits_{s>p}u_{sp}n_sn_p=\sum\limits_sn_s(\epsilon _s+\tilde
\epsilon _s)/2=E 
\end{equation}
Note, that the sums in (\ref{eqs} , \ref{eps}) containing the residual
interaction $u_{sp}$ can be substantially reduced by a proper choice of the
mean field basis (for instance, the terms $u_{sp}$ can have different signs
in such a basis). In practice, the values $\epsilon _s$ and $\tilde \epsilon
_s$ may be very close. Since in the above expressions (\ref{eqs}) the
nondiagonal matrix elements of the interaction are not taken into account,
one can expect that the distribution of occupation numbers defined by these
equations gives a correct result if the interaction is weak enough (ideal
gas approximation). However, below it will be shown that, in fact, even for
strong interaction the Fermi-Dirac distribution can be also valid if the
total energy $E$ is rescaled in a proper way, by taking into account the
increase of the temperature due to statistical effects of the (random)
interaction.

One should also note that the above consideration is similar to the standard
derivation (see e.g. \cite{BM69}) of the Fermi-Dirac distribution from the
canonical distribution (\ref{gibbs}) for the case of many non-interacting
particles (ideal gas). It is curious that the Fermi--Dirac distribution is
very close to the canonical distribution (\ref{gibbs}) even for very small
number of particles, $n\geq 2,$ provided the number of essentially occupied
orbitals is large (which happens for $T\gg \epsilon $ or $\mu \gg \epsilon $
). This fact results from the large number of ``principal'' terms in the
partition function $Z_s\,$ , and allows one to replace $\alpha _s$ by $%
\alpha $ in the term $Z_s(n,T)/Z_s(n-1,T)\equiv \exp (\alpha _s+\beta T)$ in
the canonical distribution (\ref{gibbs}) (compare with (\ref{nsZ})).

One should stress, however, that the temperature $T\,$ in the Fermi-Dirac
distribution is different from that in the canonical distribution. Indeed,
using the expansion $\alpha _s=\alpha (\epsilon _F)+\alpha ^{\prime
}(\epsilon _s-\epsilon _F)$ one can obtain the relation between the
Fermi-Dirac ($\beta _{FD})$ and canonical ($\beta )$ inverse temperatures, $%
\beta _{FD}=\beta +\alpha ^{\prime }\epsilon _F$ . Concerning the chemical
potential, its definition also changes, $-\mu /T=\alpha (\epsilon _F)-\alpha
^{\prime }\epsilon _F$ . More specifically, for the same total energy $E$ of
the system , the canonical and Fermi-Dirac distributions give the same
distribution $n_s$ defined, however, by different temperatures, see details
in \cite{FIC96,FGI96,FI97} and discussion below.

\subsection{Analytical approach to the $n_s-$distribution}

In the previous section it was shown how the standard Fermi-Dirac
distribution occurs in the TBRI-model when number of particles is very
large. However, the expression (\ref{ns}) for the distribution of occupation
numbers via the shape of chaotic eigenstates is of more general form and
also valid even when the number of particles is relatively small. In this
case the $n_s-$distribution can be of the form very different from the
FD-distribution. Below we show how to analytically derive the $n_s-$%
distribution and express it in terms of single-particle and unperturbed
many-particle spectrum, using general properties of the $F-$function \cite
{FI97}.

For simplicity, we consider the case of relatively strong interaction, when
the shape of the LDOS and exact eigenstates can be described by the
Gaussian. In order to calculate the occupation numbers $n_s$, we use the
expression (\ref{nsZ}) containing two partial partition functions $Z_s(n,E)$
and $Z_s(n-1,E-\epsilon _s)$ which correspond to systems with $n$ and $n-1$
particles, with the orbital $s\,$ is excluded from the set of
single-particle states. The partition function can be found from the
relation 
\begin{equation}
\label{Z}Z(E)=\sum_k\tilde F(E_k-E)\approx \int \rho _0(E_k)\tilde
F(E_k-E)dE_k 
\end{equation}
As was discussed above, the density of unperturbed states $\rho _0(E_k)$ in
the TBRI-model it the Gaussian, s 
\begin{equation}
\label{rho0}\rho _0(E_k)=\frac N{\sqrt{2\pi \sigma _0^2}}\exp \left( -\frac{%
\left( E_k-E_c\right) ^2}{2\sigma _0^2}\right) 
\end{equation}
where $E_c$ is the center of the energy spectrum and $N$ is the total number
of states. If the shape of eigenstates is also described by the Gaussian, 
\begin{equation}
\label{Fgauss}\tilde F(E_k-E)=\frac N{\sqrt{2\pi (\Delta E)^2}}\exp \left( - 
\frac{\left( E_k-E\right) ^2}{2(\Delta E)^2}\right) 
\end{equation}
then the integration in (\ref{Z}) can be easily performed. The variance $%
(\Delta E)^2$ is defined by Eq.(\ref{H2}). It should be pointed out that,
strictly speaking, in this expression the center of the $F-$function is
shifted by the value $\Delta _1^{(i)}$ from the unperturbed energy, $%
E=E^{(i)}+\Delta _1^{(i)}$ , see details in \cite{FI97}. This shift is due
to the level repulsion which forces eigenvalues $E^{(i)}$ in the lower part
of the spectrum to move down. The mean-field energies $E_k=H_{kk}$ do not
include the nondiagonal interaction which results in the repulsion.
Therefore, the ``center '' of the $F-$function is shifted by the value $%
\Delta _1^{(i)}=H_{ii}-E^{(i)}$ . This shift is estimated in Ref. \cite{FI97}
as follows, 
\begin{equation}
\label{delta1i}\Delta _1^{(i)}\approx \left( E_c-E^{(i)}\right) \frac{%
(\Delta E)^2}{2\sigma _0^2} 
\end{equation}
where $\sigma _0^2$ is the variance of the unperturbed spectrum. One can see
that since the variance $(\Delta E)^2\,$of the LDOS is typically much
smaller than $\sigma _0^2$ , this shift in many cases can be neglected.

Direct integration in Eq.(\ref{Z}) gets 
\begin{equation}
\label{ZE}Z(E)=\frac N{\sqrt{2\pi \sigma ^2}}\exp \left( -\frac{\left(
E-E_c\right) ^2}{2\sigma ^2}\right) 
\end{equation}
where $\sigma ^2=\sigma _0^2+(\Delta E)^2$ , therefore, the variance of the
partition function $Z(E)\,$coincides with the variance of the perturbed
spectrum. In order to calculate the occupation numbers $n_s$, one should use
the expression (\ref{nsZ}). Therefore, one needs to find the partition
functions $Z_s(n,E)$ and $Z_s(n-1,E-\epsilon _s)$ corresponding to $n$ and $%
n-1$ particles, with the orbital $s\,$ excluded from single-particle
spectrum. To do this, one needs to calculate the number of states $N_s$ and
the center $E_{cs}$ for these truncated systems, 
$$
N_s(n,m-1)=\frac{(m-1)!}{(m-1-n)!\,n!};\,\,\,\,\,\,\,\,\,\,\,\,\,%
\,N_s(n-1,m-1)=\frac{(m-1)!}{(m-n)!\,(n-1)!} 
$$
\begin{equation}
\label{single}E_{cs}(n)=\overline{\epsilon _{-s}}\,n\,;\,\,\,\,%
\,E_{cs}(n-1)=(\overline{\epsilon _{-s}})(\,n-1);\,\,\,\,\,\,\,\overline{%
\epsilon _{-s}}=\frac{\sum_{p\neq s}\epsilon _p}{m-1} 
\end{equation}
The variance $\sigma _{0s}$ of the energy distributions can be estimated as%
$$
\sigma _{0s}^2(n)\approx \sigma _{1s}^2\,n\,\,\,;\,\,\,\sigma
_{0s}^2(n-1)\approx (\sigma _{1s}^2)\,(n\,-1)\,\,\, 
$$
where $\sigma _{1s}^2$ is the variance of single-particle spectrum with the
excluded orbital $s$ . Here, for simplicity, we have neglected the Pauli
principle which is valid for $m\gg n$ . Finally, the distribution of
occupation numbers takes the form%
$$
n_s(E)=\frac 1{1+R} 
$$
\begin{equation}
\label{R}R=\frac{m-n}n\frac{\sigma _s(n-1)}{\sigma _s(n)}\exp \left[ -\frac{%
\left( E-E_{cs}(n)\right) ^2}{2\sigma _s^2(n)}+\frac{\left( E-\epsilon
_s-E_{cs}(n-1)\right) ^2}{2\sigma _s^2(n-1)}\right] 
\end{equation}
where $\sigma _s^2=\sigma _{s0}^2+(\Delta E)^2$ . Numerical data for the
TBRI-model are presented in Fig.16 from which very good agreement with (\ref
{R}) is seen.

\begin{figure}[htb]
\vspace{-2.0cm}
\begin{center}
\hspace{-2.9cm}
\epsfig{file=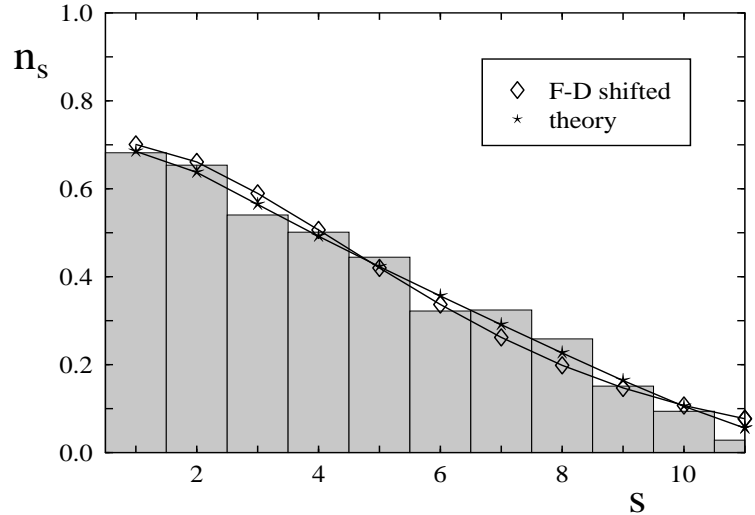,width=4.4in,height=3.6in,angle=-90}
\vspace{-1.8cm}
\caption{
Analytical description of the occupation numbers. The data are given
for the TBRI-model for the parameters of Figs.1-5 $(n=4,$ $%
m=11,V_0=0.12,d_0=1).$ The histogram is obtained according to (\ref{ns}) by
the averaging over eigenstates with energies taken from a small energy
window centered at $E=17.33$ and over $20$ Hamiltonian matrices (\ref{H})
with different realization of the two-body random interaction. Stars 
represent the analytical expression (\ref{R}) with $\sigma _{0s}$ found from
the single-particle energy spectrum. Diamonds correspond to the Fermi-Dirac
distribution with renormalized energy, see Section 3.5.
}
\end{center}
\end{figure}

It is instructive to compare this result with the Fermi-Dirac distribution
which is valid for large number of particles. In this case $R=\exp
((\epsilon _s-\mu )/T_{th})$ where $T_{th}=\sigma ^2/(E_c-E)$ is the
thermodynamic temperature which is discussed below, see (\ref{Ttherm}). The
chemical potential $\mu $ can be found numerically from the condition of
fixed total number of particles $n$.

\subsection{Effective Fermi-Dirac distribution for finite systems}

In previous Section the distribution of occupation numbers has been derived
without any reference to the temperature, from the $F-$function and
properties of the unperturbed system. However, the $n_s-$distribution in
Fig.16 seems to have a Fermi-Dirac form. One should remind that the latter
form in conventional statistical mechanics can be derived for ideal gas of
very large {\it non-interacting} particles. In such a derivation, the
presence of the thermostat is assumed, actually, in order to have
statistical equilibrium in the system. Indeed, for an isolated systems with
large $n\rightarrow \infty $ , any extremely weak interaction with an
environment results in strong statistical properties of a system. Using
modern language, one can speak about the {\it onset of chaos} due to this
interaction. In fact, the weakness of the coupling to the heat bath gives
the possibility to treat the gas of particles as an ideal gas. It is well
known, that in this case one can write the following equations, 
\begin{equation}
\label{standFD}\sum_sn_s=n,\,\,\,\,\,\,\,\,\,\,\,\,\,\,\sum_s\epsilon
_sn_s=E 
\end{equation}
where $n$ and $E\,$ are total number of particles and total energy, and $n_s$
is assumed to have Fermi-Dirac form, 
\begin{equation}
\label{FDform}n_s=\frac 1{1+\exp \left( \alpha +\beta \epsilon _s\right) } 
\end{equation}

When in an isolated system described by the TBRI-model, the number of
particles is very large, the above equations result in the FD-distribution,
see Section 3.3. However, in such a case the interaction $V_0$ has to be
very weak. Now, if we consider the model with finite and not large number of
particles, for a weak interaction there is no chaos in the sense that exact
eigenstates have small number of principal components $N_{pc}\ll 1$ .
Therefore, this model does not allow for its statistical description, in
other words, there is no statistical equilibrium. For example, if for such a
case we compute the $n_s-$distribution according 
to the definition (\ref{ns}),
there are very large fluctuations in occupation numbers when slightly
changing the total energy $E^{(i)}$ , see Fig.17.

\begin{figure}[htb]
\vspace{-2.0cm}
\begin{center}
\hspace{-2.9cm}
\epsfig{file=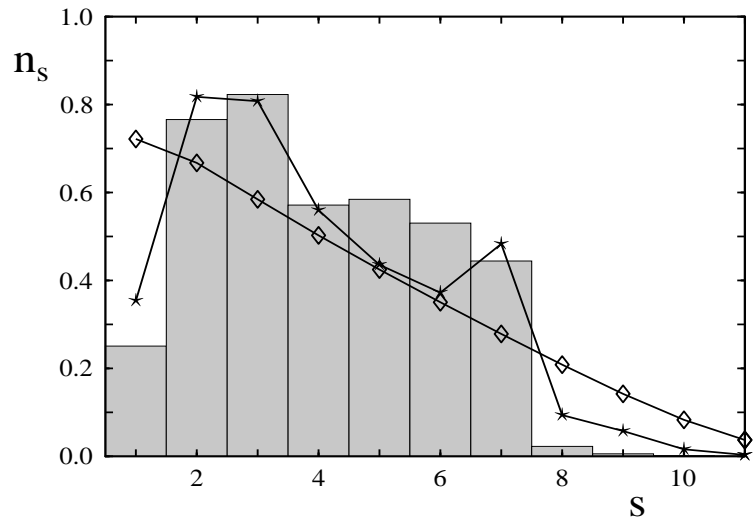,width=4.4in,height=3.6in,angle=-90}
\vspace{-1.8cm}
\caption{
Absence of a statistical equilibrium for the distribution of the
occupation numbers. The histogram is obtained in the same way as in Fig.16,
for a very weak interaction $V=0.02$ which correspond to the region of (II)
of the ``initial chaotization'', see Section 3.7. The total energy is 
$E^{(i)}=17.33$. Stars correspond to the theoretical expression (\ref{R})
which is not valid in this region due to absence of equilibrium. Circles are
obtained by direct numerical computation of $n_s$ with the $F-$function
taken in a specific form, and with the summation performed over real
unperturbed spectrum (instead of the integration with the Gaussian
approximation for $\rho _o$ 
used in Eq. (\ref{R})), 
see details in \protect\cite{FI97}. 
}
\end{center}
\end{figure}

In order to have chaotic eigenstates, and, as a result, the possibility of
statistical description, one needs to increase the interaction in order to
exceed the threshold $V_0\geq d_f$ (see also discussion in Section 3.7). The
less number of particles, the stronger interaction is needed since the
two-particle density $\rho _f=d_f^{-1}$ strongly depends on the number of
particles. On the other hand, if interaction is strong, the second equation
in Eq.(\ref{standFD}) is not correct and can not be used for the derivation
of the $n_s-$distribution. To demonstrate this, in Ref.\cite{FI97} the
distribution of occupation numbers $n_s$ for the two-body random interaction
model has been directly computed according to Eq.(\ref{ns}) from exact
eigenstates of the Hamiltonian matrix (\ref{H}), see also \cite{FIC96,FI97}.
These data for the ``experimental'' values of $n_s$ are shown in Fig.18 by
the histogram which is obtained from the average over small energy window in
order to smooth the fluctuations (with additional averaging over different
realizations of the two-body random matrix elements). To compare with the
standard Fermi-Dirac distribution, Eqs.(\ref{standFD}) have been also
numerically solved in order to find the temperature and chemical potential,
the resulting $n_s-$distribution is shown by circles. One should stress that
the value of the energy $E$ in (\ref{eqs}) was taken the same as for the
exact eigenstates from which actual distribution of $n_s$ was computed,
namely, $E\approx E^{(i)}\,$. The comparison of the actual distribution
(histogram) with the ``theoretical '' one, reveals a big difference for a
chosen (quite strong) perturbation $V=0.20\,.$

To describe correctly the $n_s$-distribution in terms of the Fermi-Dirac
distribution, in Ref.\cite{FI97} it was suggested to renormalize the total
energy of the systems due to the interaction between particles, and instead
of Eq.(\ref{standFD}) to solve the following equations,

\begin{equation}
\label{newFD}\sum_sn_s=n,\,\,\,\,\,\,\,\,\,\,\,\,\,\,\sum_s\epsilon
_sn_s=E+\Delta _E 
\end{equation}
where $\Delta _E\,$ is the shift of the total energy due to the interaction.
In Ref.\cite{FI97} it was argued that in the case of random interaction,
this term absorbs statistical increase of the energy and gives the correct
result for the $n_s-$distribution. In fact, this assumption is based on a
deep equivalence between the {\it external chaos} originated by the heat
bath in the case of open systems, and {\it internal chaos} due to a random
character of the interaction. Therefore, it was assumed that random
interaction and chaos in closed systems plays the role of a heat bath.

\begin{figure}[htb]
\vspace{-2.0cm}
\begin{center}
\hspace{-2.9cm}
\epsfig{file=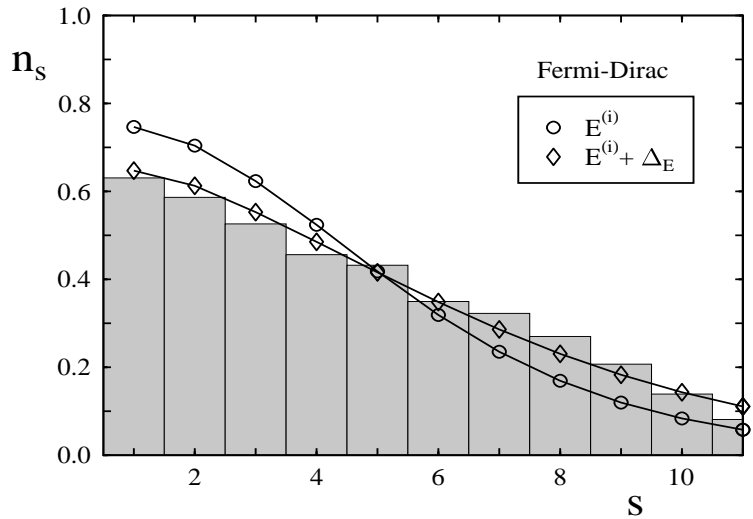,width=4.4in,height=3.6in,angle=-90}
\vspace{-1.8cm}
\caption{
Fermi-Dirac distribution with and without additional shift of energy
due to a (strong) interaction, $V_0=0.20$ (other parameters are the same as
in Fig.16). Circles stand for the Fermi-Dirac distribution obtained for the
total energy $E=17.3$ corresponding to the energy of eigenstates, see (\ref
{eqs}). Diamonds correspond to the distribution obtained for the energy
shifted according to Eq. (\ref{Delfinal}).
}
\end{center}
\end{figure}

In order to find analytically the shift $\Delta _E$ , one needs to consider
the structure of exact eigenstates in the unperturbed basis, in particular,
to find the shift between the energy of an exact eigenstates and the mean
energy of the components of the same eigenstate \cite{FI97}. Since the
density of states rapidly increases with the energy $E$ , the number of
higher basis states admixed to an eigenstate due to the interaction is
larger than the number of lower basis states. As a result, the mean energy 
\begin{equation}
\label{Eki}\left\langle E_k\right\rangle _i=\sum_kE_kF_k^{(i)}\approx \int
E_kF_k^{(i)}\rho _0(E_k)dE_k 
\end{equation}
of the components in an exact eigenstate $\left| i\right\rangle $ is higher
than the eigenvalue $E^{(i)}$ corresponding to this eigenstate (we consider
here the eigenstates in the lower part of the spectrum). There is another
effect which decreases the value of $\left\langle E_k\right\rangle _i$ , see
(\ref{Eki}), which remains even if the density of states does not depend on
the energy. This second effect is due to repulsion between energy levels,
according to which the eigenvalues move down for this part of the spectrum,
therefore, the difference between $\left\langle E_k\right\rangle _i$ and $%
E^{(i)}$ increases due to the interaction. The second effect also shifts the
``center'' of the $F-$function. One should stress that all effects leading
to the above shift of the energy are automatically taken into account in the
relation (\ref{Eki}). Thus, one can analytically calculate this shift $%
\Delta _E=\left\langle E_k\right\rangle _i-E^{(i)}$ from the equation (\ref
{Eki}). For this, one needs to know the unperturbed density of states and
the form of $F-$function. The evaluation of the shift $\Delta _E$ has been
done in \cite{FI97} by assuming some generic form for the $F-$ function
which is valid in a wide range of the interaction strength $V$ ,

\begin{equation}
\label{Delfinal}\Delta _E=\left\langle E_k\right\rangle _i-E^{(i)}=\frac{%
\left( \Delta E\right) ^2}{\sigma _0^2}\left( E_c-E\right) 
\end{equation}
where $E_c$ is the center of the energy spectrum.

Thus, to find correct values for the occupation numbers in the Fermi-Dirac
distribution, we should solve Eqs.(\ref{newFD}) with $\Delta _E\,$ defined
by Eq.(\ref{Delfinal}). The resulting $n_s-$dependence is shown in Fig.18 by
diamonds. As one can see, such a correction gives quite good correspondence
to the numerical data. Similar difference occurs for larger number of
particle (and smaller interaction strength), see Fig.19 where due to serious
numerical problems, the data for the occupation number distribution are
given without direct comparison with actual $n_s-$distribution.

\begin{figure}[htb]
\vspace{-2.0cm}
\begin{center}
\hspace{-2.9cm}
\epsfig{file=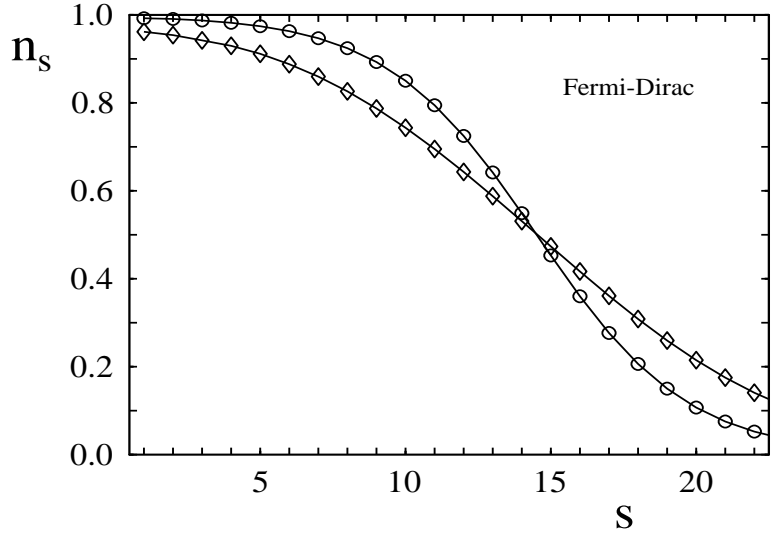,width=4.4in,height=3.6in,angle=-90}
\vspace{-1.8cm}
\caption{
The same as in Fig.18 for large number of particles, $n=14,$ and
orbitals, $m=28$ , for $V_0=3\cdot 10^{-3}$ and $d_0=1$ . The histogram is
not possible to obtain numerically in the direct computation, due to a very
large size of the Hamiltonian matrix, $N=330000$ .
}
\end{center}
\end{figure}

To check the analytical prediction (\ref{Delfinal}) for the shift $\Delta _E$
, this shift has been directly calculated by comparing the energy $E^{(i)}$
of exact eigenstates with the energy $\left\langle E_k\right\rangle _i$ .
The latter has been numerically found from the exact relation $\left\langle
E_k\right\rangle _i=\sum_kE_k\left| C_k^{(i)}\right| ^2$ (compare with (\ref
{Eki})). The comparison of these data (circles in Fig.20) with Eq. (\ref
{Delfinal}) (straight full line) shows reasonable agreement, if to neglect
strong fluctuations around the global dependence. These fluctuations are due
to fluctuations in the components of specific exact eigenstates $\left|
i\right\rangle $ (note, that the presented data correspond to individual
eigenstates, without any additional averaging).

\begin{figure}[htb]
\vspace{-3.5cm}
\begin{center}
\hspace{-1.6cm}
\epsfig{file=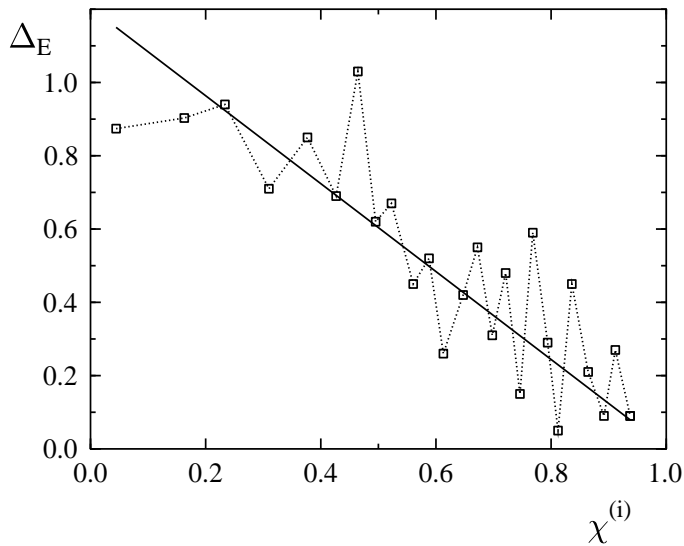,width=4.4in,height=3.6in,angle=-90}
\vspace{-1.0cm}
\caption{
Shift of the total energy for the corrected Fermi-Dirac
distribution. The data are given for the TBRI-model with $%
n=4,m=11,d_0=1,V=0.12\,$. The straight line is the analytical expression (%
\ref{Delfinal}), the dotted line (squares) presents direct computation of
the shift based on the diagonalization of the Hamiltonian (\ref{H}) with the
following computation of the $<E_k>_i$ . On the horizontal axes the rescaled
energy $\chi ^{(i)}=(E^{(i)}-E_{fermi})/(E_c-E_{fermi})$ is plotted.
}
\end{center}
\end{figure}

\subsection{Temperature vs. chaos}

In previous Section it was shown that if the $n_s-$distribution for finite
number of particles in the TBRI-model has the from of the Fermi-Dirac
distribution, it can be found from the modified Eq.(\ref{newFD}). These
equations also determine the parameters $\alpha \,$ and $T=\beta ^{-1}\,$
which may be associated with the ``{\it chemical potential} '' and ``{\it %
temperature} '' for an isolated system. One can see, that due to the shift $%
\Delta _E$ of the total energy in Eq.(\ref{newFD}), there is a corresponding
shift of the temperature $\Delta T\,$ . It is important to note that in
systems with infinite number of particles, all definitions of temperature
give the same result. In contrast to this, for finite number of particles
different definitions give different results.

Let us, first, start with the standard definition of temperature,

\begin{equation}
\label{Ttnew}\frac 1{T_{th}}=\frac{dS_{th}}{dE}=\frac{d\ln \,\rho }{dE} 
\end{equation}
where $S_{th}$ is the thermodynamical entropy,

\begin{equation}
\label{St}S_{th}=\ln \,\rho (E)\,+\,const 
\end{equation}
and $\rho (E)$ is the density of states.

In fact, this definition of the {\it thermodynamical temperature} stems from
the estimate of the position of maximum of the canonical averaging function $%
\Phi _T(E)$ , see (\ref{fi}), if we assume that the position of its maximum $%
E_m$ coincides with the energy $E$ of a system. One should stress that in
the above definition $\rho (E)$ is the total density of states, therefore,
the interaction is essentially taken into account.

Another definition, which is consistent with the first law of thermodynamics
(energy conservation), is given by the relation $\left\langle E\right\rangle
_T=E$ , see Eq.(\ref{energy}). Here the averaging is performed over the
canonical distribution (\ref{gibbs}). Since the width $\Delta _T$ of the
canonical averaging function $\Phi _T(E)$ is not zero, the two definitions
of the temperature, (\ref{Ttnew}) and (\ref{energy}) give, in principal,
different values. Indeed, in the case of the Gaussian form of $\rho (E)$ the
value of $T_{th}$ given by (\ref{Ttnew}) takes the form (see also \cite
{HZB95,ZBHF96}), 
\begin{equation}
\label{Ttherm}T_{th}=\frac{\sigma ^2}{E_c-E} 
\end{equation}
where $E_c$ and $\sigma $ are the center and the width of the total density $%
\rho (E).$

On the other hand, direct evaluation of the relation (\ref{energy}) leads to
the following definition of the {\it canonical temperature}, 
\begin{equation}
\label{Tcan}T_{can}=\frac{\sigma ^2}{E_c-E+\Delta } 
\end{equation}
Here, the shift $\Delta $ is approximately given by the expression 
\begin{equation}
\label{DeltaT}\Delta =\frac \sigma K\left[ \exp \left( -\frac{\left( E_{\min
}-E_m\right) ^2}{2\sigma ^2}\right) -\exp \left( -\frac{\left( E_{\max
}-E_m\right) ^2}{2\sigma ^2}\right) \right] 
\end{equation}
where 
\begin{equation}
\label{K}K=\int\limits_{x_{\min }}^{x_{\max }}\exp \left( -\frac{x^2}%
2\right) dx\approx \sqrt{2\pi }\,\,\,;\,\,\,\,x=\frac{E-E_m}\sigma
\,\,;\,\,\,E_m=E_c-\frac{\sigma ^2}{T_{can}} 
\end{equation}
One can see that the shift $\Delta $ itself depends on the temperature, it
is proportional to the width $\Delta _T=$$\sigma $ of the function $\Phi
_T(E)$. In the above expressions, $E_{\min }$ and $E_{\max }$ are the low
and upper edges of the energy spectrum. Note that the relation $\Delta =0$
occurs at the center of the spectrum, therefore, at the center the
temperature diverges and in the upper part of the spectrum is negative. This
fact is typical for systems with bounded spectrum, for example, for spin
systems. In fact, the TBRI-model (\ref{H}) with finite number $m$ of
orbitals can be treated as a model for one open shell in atoms, nuclei,
clusters, etc. However, in realistic many-body systems there are always
higher shells which contribute to the density of states for higher energy.
Thus, the density of states $\rho (E)$ is a monotonic function which results
in the positive temperature. For such physical applications, the model (\ref
{H}) with finite number of orbitals is reasonable in the lower part of the
energy spectrum where the influence of higher shells can be neglected.

\begin{figure}[htb]
\vspace{-2.3cm}
\begin{center}
\hspace{-2.1cm}
\epsfig{file=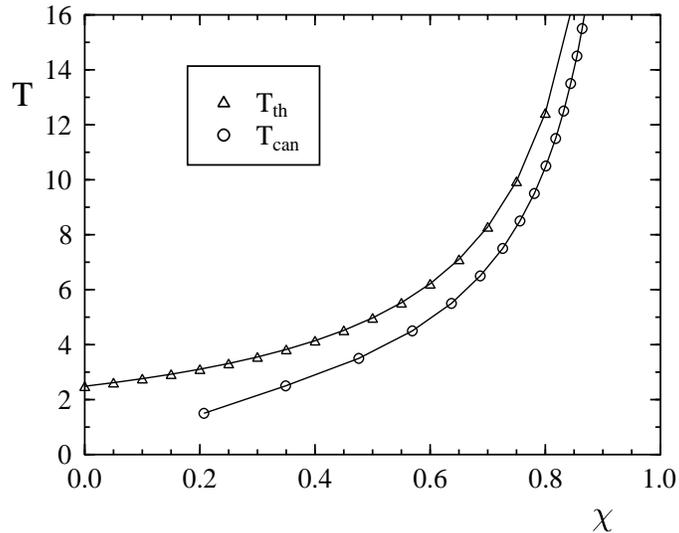,width=4.4in,height=3.6in,angle=-90}
\vspace{-1.5cm}
\caption{
Different temperatures versus the rescaled energy $\chi
=(E-E_{fermi})/(E_c-E_{fermi})$ for the TBRI-model . Parameters are the same
as in Fig.16. Triangles stand for the thermodynamical temperature $T_{th}$
defined by (\ref{Ttherm}) and should be compared to the canonical
temperature $T_{can}$ (circles), see (\ref{Tcan}). The width $\sigma $ of
the perturbed density of states is defined by the residual interaction $%
V_0=0.12$ according to (\ref{H2}) and the relation $\sigma ^2=\sigma
_0^2+(\Delta E)^2$ with $\sigma _0$ found numerically from the unperturbed
many-particle energy spectrum.
}
\end{center}
\end{figure}

One can also see that the difference between the two equations of state $%
T(E) $ defined by Eqs.(\ref{Ttherm}) and (\ref{Tcan}), disappears for highly
excited eigenstates (for which $E_m-E_{\min }\gg \sigma ),$ or in large
systems with $n\gg 1$ . Indeed, one can obtain, $E_c-E\sim n\sigma _1$ ,
where $\sigma _1$ is the width of single-particle spectrum. On the other
hand, according to the central limit theorem, the variance of total energy
spectrum can be estimated as $\sigma _0^2=\sum_n\sigma _1^2\approx n\sigma
_1^2$ , therefore, the ratio $\sigma /(E_c-E)\sim 1/\sqrt{n}$ tends to zero
at $n\rightarrow \infty $ . As was mentioned above, in such realistic finite
systems like atoms and nuclei, the number of particles in an open shell is
relatively small ($n=4$ for the Ce atom \cite{FGGK94} and $n=12$ in nuclear
shell model \cite{HZB95,ZBHF96}), therefore, the corrections to the
thermodynamical temperature (\ref{Ttnew}) may be important, especially, for
low energies. The detailed discussion of different temperatures in nuclear
shell model is given in \cite{HZB95,ZBHF96}.

The energy dependence of temperatures $T_{th}$ and $T_{can}$ , as well as
the temperature $T_{\exp }$ found directly from the numerical simulation, is
shown in Fig.21. The data refer the TBRI-model with $n=4$ interacting
Fermi--particles and $m=11$ orbitals. The comparison of the thermodynamical
temperature $T_{th}$ defined by (\ref{Ttherm}), with the ``canonical''
temperature (\ref{Tcan}) reveals quite strong difference in all the range of
the rescaled energy $\chi =(E-E_{fermi})/(E_c-E_{fermi})$.

Now let us find the shift of the temperature $\Delta T\,$ which is due to
the interaction, see previous Section. Since it is directly related to the
shift of total energy, $E\equiv \left\langle E_k\right\rangle
_i=E^{(i)}+\Delta _E$ , one can get, 
\begin{equation}
\label{Tfinal}T=T_0+\Delta T=\frac{\sigma _0^2}{E_c-E^{(i)}-\Delta _E}%
\approx \frac{\sigma _0^2}{E_c-E^{(i)}}+\frac{\left( \Delta E\right) ^2}{%
E_c-E^{(i)}} 
\end{equation}
Therefore, statistical effects of random interaction can be directly related
to the increase of temperature of a system, $\Delta T/T_0=(\Delta
E)^2/\sigma _0^2$ .

One of the important questions deals with thermodynamical description of
isolated systems of interacting particles. In any thermodynamical approach
one needs to define, in a consistent way, such quantities as entropy,
temperature and equation of state. This problem has been recently discussed 
\cite{HZB95,ZBHF96} in application to shell models of heavy nuclei. In
particular, it was shown that for a realistic residual interaction different
definitions of temperatures give the same result.

\subsection{Transition to chaos and statistical equilibrium}

Let us now summarize the discussed above results for the TBRI-model in what
concerns transition to chaos and thermalization. The latter term is not
defined for isolated systems, our suggestion is to treat ``{\it %
thermalization} '' as the onset of {\it statistical equilibrium}. The latter
allows to give a reasonable statistical description and may be used to find
thermodynamical description.

Depending on the strength of (two-body) interaction between particles , one
can fix different situations in the model. First region (I) refers {\it %
strong (perturbative) localization.} This occurs when the interaction is
very weak, $V_0\ll d_f$ and standard perturbation theory gives correct
result. In this case exact eigenstates have only few relatively large
components ($N_{pc}\sim 1)$ , in other words, the eigenstates are strongly
localized in the unperturbed basis. This situation is quite typical for
lowest eigenstates (where the density of states is small) even if for higher
energies the eigenstates can be considered as very ``chaotic'' ones ($%
N_{pc}\gg 1)$ .

The second region (II) is characterized by an {\it initial chaotization} of
exact eigenstates which corresponds to a relatively large, $N_{pc}\gg 1$
number of principal components and $V<d_f$. The latter condition is
essential since it results in very strong (non-Gaussian) fluctuations of
components $C_k^{(i)}$ \cite{FGS97} for the fixed energy $E^{(i)}$ of
compound state $\left| i\right\rangle \,.$ Such a type of fluctuations
reflects itself in a specific character of eigenstates, namely, they turn
out to be {\it sparsed}. As a result, the number of principal components can
not be estimates as $N_{pc}\approx \Gamma /D$ , as is typically assumed in
the literature. In this case the energy width $\Gamma $ of both the LDOS\
and eigenstates can be approximately describe as $\Gamma \approx 2\pi
V_0^2/d_f$ . In Fig.17 it is shown how the distribution of occupation
numbers $n_s$ looks like for the TBRI-model (\ref{H}) with $n=4$ particles
and $m=11$ orbitals and very weak perturbation $V/d_0\approx 0.02$ . One can
see that the distribution of occupation numbers has nothing to do with the
Fermi-Dirac distribution (full diamonds), it turns out to be even the
non-monotonic function of the energy $\epsilon _s$ of orbitals (see also 
\cite{FIC96}). Note that the averaging procedure used in Fig.17 can not wash
out strong fluctuations in occupation numbers $n_s$ .

With further increase of the interaction, where $N_{pc}\gg 1$ and $V>d_f\,$
, the region (III) of the {\it statistical equilibrium} emerges. In this
region the fluctuations of eigenstate components $C_k^{(i)}$ are of the
Gaussian form \cite{FGS97} and one can introduce the $F-$function (\ref
{nalpha}) as the shape of exact eigenstates in the unperturbed energy basis.
Correspondingly, the fluctuations of the occupations numbers $n_s$ are small
in accordance with the central limit theorem, $\Delta n_s/n_s\sim
N_{pc}^{-1/2}\ll 1$ for $n_s\sim 1$ . One should stress that in this region
the value of $N_{pc}$ is given by the common estimate, $N_{pc}\sim \Gamma /D$
. As a result, the $n_s-$distribution changes slightly when changing the
energy of a system. Such a situation can be naturally related to the {\it %
onset of thermal equilibrium}, though the form of the distribution $n_s$ can
be quite different from the Fermi-Dirac distribution. In this case, the $F-$%
distribution allows for a correct description of an actual distribution of
occupation numbers in isolated quantum systems of interacting particles. One
can see that the equilibrium distribution for the occupation numbers arises
for much weaker condition compared to that needed for the Fermi-Dirac
distribution. Since the energy interval $d_f$ between directly coupled basis
states is small, it is enough to have a relatively weak residual interaction 
$V>d_f$ in order to have the equilibrium distribution (note, that the value
of $d_f$ decreases rapidly with the excitation energy).

Next region (IV) is where the {\it canonical distribution} (\ref{gibbs})
occurs; for this case in addition to the equilibrium, one needs to have
large number of particles, $n\gg 1.$ If, also, the condition $\Gamma \ll
nd_0 $ is fulfilled, the standard Fermi-Dirac distribution is valid with a
proper shift of the total energy due to the interaction, see Section 3.5.
Typically, this region is associated with the onset of the canonical
thermalization (see, for example, \cite{HZB95,ZBHF96}). In practice,
the condition (IV) of the canonical thermalization is not easy to satisfy in
realistic systems like atoms or nuclei since $n$ in the above estimates is,
in fact, the number of ``active'' particles (number of particles in a
valence shell) rather than the total number of particles. Thus, the
description based on the $F-$distribution (\ref{nalpha}) which does not
require the canonical thermalization condition (IV), is more accurate.

The above statements are confirmed by the direct numerical study of the
two-body random interaction model \cite{FIC96,FI97} with few particles when
changing the interaction strength $V/d_0.$ If, instead, we increase the
number of particles keeping the interaction small, $V\ll d_0$ , the
distribution (\ref{nalpha}) tends to the Fermi-Dirac one as it is expected
for the ideal gas, see \cite{FI97}.

\section{ Concluding remarks}

In this paper we have discussed a novel approach to isolated systems of
finite number of interacting particles. The goal of this approach is a
direct relation between the average shape of exact eigenstates ($F-$%
function), and the distribution $n_s\,$of occupation numbers of
single-particle levels. From this relation one can see that there is no need
to know exactly the eigenstates, instead, if these eigenstates are chaotic (
random superposition of a very large number of components of basis states),
the $F-$function absorbs statistical effects of interaction between
particles and determines the form of the $n_s-$distribution. Therefore, the
structure of chaotic eigenstates in dependence on the model parameters, is
the central question in this approach.

The results discussed above relate to the TBRI-model for which all two-body
matrix elements are assumed to be random and independent variables. This
assumption was made in order to study limiting statistical properties of
this model. In particular, it was shown that even in this limit case of
completely random (two-body) interaction, the Hamiltonian matrix in
many-particle representation can not be treated as the random matrix,
therefore, the RMT is, strictly speaking, not valid. However, under some
conditions exact eigenstates turn out to be quite random and statistical
approach is correct, however, one should take into account the form of
eigenstates in a given basis of unperturbed part $H_0$ .

Another reason for the study of this TBRI-model, is that it is very
convenient for the demonstration of the developed approach. In this case
there are no any effects of regular motion in the system, and many of
analytical estimates can be obtained in a clear way. In particular, it was
shown how to calculate the $n_s-$ distribution from the shape of
eigenstates, provided the unperturbed spectrum of energy is known. One can
stress that in this way one can analytically obtain the $n_s-$distribution
which can have the form very different from the standard Fermi-Dirac
distribution. On the other hand, for sufficiently large interaction the $%
n_s- $distribution is of the Fermi-Dirac form, therefore, it is convenient
to introduce an effective ``temperature '' and ``chemical potential '' which
give correct description of actual $n_s-$distribution in terms of the
Fermi-Dirac distribution. To do this, one needs to find the shift of the
total energy which formally comes into equations determining the
FD-distribution. One should stress that this shift is directly related to
the $F-$function and can be found analytically.

Now, we would like to point out that the approach discussed in this paper
for the TBRI-model is of generic and can be applied both for random and
dynamical interaction. One of the most interesting problems is the
application of the approach to dynamical systems with the well-defined
classical limit. In this situation exact eigenstates in the corresponding
quantum model can be expected to appear when two conditions are fulfilled.
The first one is the strong chaos in the classical counterpart, and the
second is the semi-classical limit (which is typically equivalent to a high
energy of a system). Under these two conditions, eigenstates of quantum
model have many components and these components may be treated as
pseudo-random, thus leading to a statistical equilibrium in the system and
possibility to apply the suggested approach.

In Refs.\cite{WIC98,BGIC98,BGI98} 
two quantum dynamical systems have been studied
and compared to their classical limits. In both cases the main results refer
the region of parameters where the classical motion is strongly chaotic. One
of the important questions which was under close investigation is the
quantum-classical correspondence for the $F-$function and the LDOS. As was
pointed out in Ref.\cite{CCGI96}, 
there is a quite clear and easy way for
finding {\it classical $F-$function} and {\it classical LDOS. }It is
instructive to explain this approach since it is of generic and can be used
in many physical applications (see details in \cite{WIC98,BGIC98,BGI98}).

Let us start with the classical $F-$function. We assume that the total
Hamiltonian can be represented in the form, 
\begin{equation}
\label{HamT}H=\,\,H_0+\,V\,\,\,;\,\,\,\,\,\,\,H_0=\sum_{k=1}^nH_k^{\,0}\,;\,%
\,\,\,\,\,\,H_k^{\,0}=H^{\,0}(p_k,x_k) 
\end{equation}
Here $H_0$ stands for the ``unperturbed '' Hamiltonian which is the sum of
partial Hamiltonians $H_k$ describing the motion of different
(non-interacting) $n$ particles. The interaction between particles is
embedded in $V\,$which is assumed to result in chaotic behavior of the
(total) system. Note that the same consideration is valid if instead of
particles we mean different degrees of freedom for one particle. Now let us
fix the total energy $E\,$ of the Hamiltonian $H(t)$ and find (numerically)
the trajectory $p_k(t)\,,\,x_k(t)$ by computing Hamiltonian equations. Since
the total Hamiltonian is chaotic, there is no problem with the choice of
initial conditions $p_k(0)\,,\,x_k(0)$ , any choice gives the same result if
one computes for sufficiently large time. When time is running, let us
collect the values of {\it unperturbed }Hamiltonian $H_0(t)$ for fixed
values $t=T,2T,3T,\,...\,,$ and construct the distribution of energies $%
E_0(t)$ along the (chaotic) trajectory of the {\it total} Hamiltonian $H$ .
In such a way, one can get some distribution $W(E_0;E=const)\,\,$ .
Comparing with the quantum model, one can see that this function $%
W(E_0;E=const)\,$ is the classical analog of the $F-$function which is the
average shape of eigenstates in energy representation. Indeed , any of exact
eigenstates corresponds to a fixed total energy $E=const$ and it is
represented in the unperturbed basis of $H_0$ , in fact, $F-$function is the
(average) projection of exact eigenstate onto the set of unperturbed ones.
Thus, one can expect that for chaotic eigenstates in a deep semiclassical
region the two above quantities, classical and quantum ones, correspond to
each other.

On the other hand, one can consider the complimentary situation. Let us fix
the {\it unperturbed} energy $E_0$ and compute a trajectory $%
p_k^{(0)}(t)\,,\,x_k^{(0)}(t)$ which belongs to the unperturbed Hamiltonian $%
H_0(t)\,.\,$ Similar to the previous case, let us put this unperturbed
trajectory into the total Hamiltonian $H(t)$ and collect the values of total
energy $E(t)$ along the unperturbed trajectory for discrete values of time.
In this way one can find the distribution $\tilde W(E;E_0=const)$ which now
should be compared with the LDOS in the corresponding quantum model.
However, in this case one should be careful and make an average over many
initial conditions $p_k(0)\,,\,x_k(0)$ with the same energy $E_{0}$
, if the unperturbed Hamiltonian is regular. In fact, the above two
classical distributions $W(E_0;E=const)$ and $\tilde W(E;E_0=const)$
determine the ergodic measure of energy shells, the first one, when
projecting the phase space surface of $H\,$ onto $H_0$ , and in the second
one, the surface of $H_0\,$ onto $H$ (see discussion in \cite{CCGI96}).

Numerical data for two different dynamical models have shown amazingly good
correspondence between the $F-$function and the LDOS, and their classical
counterparts, see details in Refs.\cite{WIC98,BGIC98,BGI98}. 
Recently, few other
systems have been studied both in classical and quantum representations, and
again, very good correspondence has been numerically found in the
semiclassical region. These data confirm the theoretical predictions, and
seems to be very important in view of future developments of the
semiclassical theory of dynamical chaotic systems. It is important to point
out that the above correspondence opens a new way for the {\it semi-quantal}
description of quantum system, when the form of the $F-$function is taken
from a classical model and used in order to find distribution of occupation
numbers $n_s\,$of single-particle levels in the corresponding quantum system 
\cite{BI99}.

Another interesting problem which has been studied in dynamical systems, is
the distribution of occupation numbers and the possibility of its analytical
description in the same way as it was done for the TBRI-model (see details
in \cite{BGI98}). One of the most interesting results obtained numerically,
is that the canonical distribution occurs in an isolated (dynamical) system
of only two interacting spin-particles, if one randomize the {\it non-zero
elements} of the interaction $V$ as close as possible to the dynamical
constrains of the model. This means that random interaction indeed plays a
role of the heat bath and allows to use statistical and thermodynamical
description for isolated systems.

\section{Acknowledgments}

I am very grateful to my co-authors F.Borgonovi, G.Casati, B.V.Chirikov,
V.V.Flambaum, Y.Fyodorov, G.F.Gribakin and I.Guarneri, with whom the works
have been done on the subject discussed in this review. This work was
supported by CONACyT (Mexico) Grant No. 28626-E.

\end{document}